\documentclass{aastex631}

\usepackage{graphicx}	
\usepackage{amsmath}	
\usepackage{amssymb}	
\usepackage{xcolor}
\usepackage{soul}
\usepackage{float}
\usepackage{lmodern}

\usepackage{dcolumn} 
\usepackage{placeins}   
\usepackage{xcolor}

\graphicspath{{./}{Figure/}}

\shorttitle{AASTeX v6.3.1 Sample article}
\shortauthors{Wang et al.}

\begin{document}
\title{The Steep-spectrum Radio-loud AGN Luminosity Function and Its Implications for Black Hole Growth and Star Formation}


\author[0009-0005-1617-2442]{Wenjie Wang}
\affiliation{Department of Physics, School of Physics and Electronics, Hunan Normal University, Changsha 410081, China}
\affiliation{Key Laboratory of Low Dimensional Quantum Structures and Quantum Control, Hunan Normal University, Changsha 410081, China}
\affiliation{Hunan Research Center of the Basic Discipline for Quantum Effects and Quantum Technologies, Hunan Normal University, Changsha 410081, China}

\author[0000-0001-6861-0022]{Zunli Yuan}
\affiliation{Department of Physics, School of Physics and Electronics, Hunan Normal University, Changsha 410081, China}
\affiliation{Key Laboratory of Low Dimensional Quantum Structures and Quantum Control, Hunan Normal University, Changsha 410081, China}
\affiliation{Hunan Research Center of the Basic Discipline for Quantum Effects and Quantum Technologies, Hunan Normal University, Changsha 410081, China}

\author{B. \v{S}laus}
\affiliation{Department of Physics, Faculty of Science, University of Zagreb,
Bijeni\v{c}ka cesta 32, 10000 Zagreb, Croatia}
\affiliation{Astronomical Observatory Institute, Faculty of Physics and Astronomy, Adam Mickiewicz University, ul. S\l oneczna 36, 60-286, Pozna\'{n}, Poland}

\author{Hongwei Yu}
\affiliation{Department of Physics, School of Physics and Electronics, Hunan Normal University, Changsha 410081, China}
\affiliation{Key Laboratory of Low Dimensional Quantum Structures and Quantum Control, Hunan Normal University, Changsha 410081, China}
\affiliation{Hunan Research Center of the Basic Discipline for Quantum Effects and Quantum Technologies, Hunan Normal University, Changsha 410081, China}

\author[0000-0003-2341-9755]{Yu Luo}
\affiliation{Department of Physics, School of Physics and Electronics, Hunan Normal University, Changsha 410081, China}
\affiliation{Key Laboratory of Low Dimensional Quantum Structures and Quantum Control, Hunan Normal University, Changsha 410081, China}
\affiliation{Hunan Research Center of the Basic Discipline for Quantum Effects and Quantum Technologies, Hunan Normal University, Changsha 410081, China}

\correspondingauthor{Zunli Yuan}
\email{yzl@hunnu.edu.cn}
\correspondingauthor{Hongwei Yu}
\email{hwyu@hunnu.edu.cn}

\begin{abstract}
We study the cosmic evolution of radio-loud active galactic nuclei (AGNs) using a beaming-minimized sample of 4{,}555 steep-spectrum sources compiled from the XXL survey, VLA-COSMOS, and other wide-field data sets. We constrain the rest-frame 1.4,GHz radio luminosity function (RLF) over \(0<z\lesssim5\), and model it with a luminosity-and-density evolution (LADE; DE+LE) framework. Among the tested parameterizations, Model~C is statistically preferred and provides a globally consistent description of the binned RLFs while remaining compatible with local RLF measurements and Euclidean-normalized source counts. In the fiducial solution, the LE term rises toward cosmic noon ($z\sim2$--3) and then flattens or mildly declines, whereas the DE term decreases monotonically with redshift. This combined evolution naturally reproduces the observed luminosity-dependent turnover redshift $z_{\rm peak}(L)$ (often termed ``cosmic downsizing'') without imposing \emph{a priori} distinct evolutionary laws for low- and high-power sources. We further show that the same LADE functional family calibrated for star-forming galaxies also describes radio-loud AGNs when fitted independently, enabling a unified two-component (SFG+AGN) model consistent with both the local RLF and source-count statistics. 
Finally, converting the inferred 1.4,GHz RLF to a kinetic luminosity function yields a radio-mode black hole accretion rate density (BHAD) traced by the steep-spectrum radio-loud AGN population. Its redshift dependence broadly tracks, in shape, the radio-based cosmic star formation rate density (after a conventional rescaling), with both histories peaking near $z\sim2$. The similarity suggests that the growth of this radio-loud AGN sub-population is broadly synchronized, at the population level, with the cosmic star-formation history of galaxies.
\end{abstract}


\section{Introduction}
\label{sec_intro}

Radio-loud active galactic nuclei (AGNs) inject mechanical energy through relativistic jets and are widely regarded as a key channel of AGN feedback in galaxy evolution \citep{Heckman2014}. Empirically, the link between supermassive black hole (SMBH) growth and galaxy evolution is supported by both indirect scaling relations between SMBH mass and host-galaxy properties \citep{magorrian1998, Ferrarese_Merritt2000, Gebhardt2000, Graham2011, Sani2011, Beifiori2012, McConnell_Ma2013} and direct feedback signatures such as powerful winds \citep[e.g.,][]{Nesvadba2008, Feruglio2010, Veilleux2013, Tombesi2015} and X-ray cavities \citep{Clarke1997, Rafferty2006, McNamaraNulsen2007, Fabian2012, Nawaz2014, Kolokythas2015}. Incorporating such feedback is now a cornerstone of modern galaxy-formation models \citep{Bower2006, Croton2006, Schaye2015, Croton2016, 2017MNRAS.467.4739K, 2018MNRAS.475..676S, Dave2019}. Quantifying the cosmic evolution of radio-loud AGNs is therefore essential for assessing the redshift dependence of jet-driven mechanical feedback and its contribution to SMBH growth.

A central statistical descriptor of the radio-loud AGN population is the radio luminosity function (RLF), which encodes the comoving space density as a function of radio luminosity and redshift \citep[e.g.,][]{Smolcic2009, Rigby2015, Pracy2016, Novak2018}. Combined with radio--kinetic scalings, the evolving RLF can be used to infer the cosmic history of jet kinetic output and SMBH growth in the kinetic (jet) mode \citep[e.g.,][]{2023MNRAS.523.5292K, Smolcic2017c, 2018A&A...620A.192C, 2021A&A...650A.127T}. Observationally, RLF measurements have established that luminous radio-loud AGNs evolve strongly and peak around $z\sim2$--3, whereas lower-luminosity sources evolve more mildly and peak at lower redshift, producing the luminosity-dependent peak redshift $z_{\rm peak}(L)$, consistent with the commonly discussed ``cosmic downsizing'' phenomenology \citep[e.g.,][]{Waddington2001,Clewley2004,Rigby2011,Rigby2015,Pracy2016}.

Recent progress on this luminosity-dependent evolution of the RLF has been enabled by increasingly large, multi-field radio-loud AGN compilations. In particular, \citet{2024A&A...684A..19S} assembled one of the largest samples to date and 
showed that a luminosity-dependent density evolution \citep[LDDE;][]{1983ApJ...269..352S,2003ApJ...598..886U} parameterization provides the statistically preferred description of their data among the tested models.
However, their compilation includes a non-negligible fraction of flat-spectrum radio sources.
These are preferentially jet-aligned systems, where Doppler boosting makes the compact jet core dominate over the optically-thin lobe emission,
yielding an observed flat spectrum and a boosted apparent radio luminosity that can bias the inferred high-luminosity evolution
\citep{UrryPadovani1995, 2019MNRAS.490.5300D, 2021A&A...650A.127T}.
A more robust route to the intrinsic RLF evolution is therefore to exclude flat-spectrum sources and focus on steep-spectrum objects that are less susceptible to Doppler boosting.

Beyond LDDE, an alternative phenomenological description is the luminosity-and-density evolution (LADE) framework \citep[e.g.,][]{1984ApJ...284...44C,2010MNRAS.401.2531A,2016ApJ...829...95Y}, which models the RLF evolution as the combined action of a luminosity-evolution (LE) term and a density-evolution (DE) term \citep[e.g.,][]{2016ApJ...829...95Y,yuan2017mixture}. In this decomposition, LE captures redshift-dependent shifts in the characteristic luminosity scale, while DE describes changes in the overall comoving abundance. Importantly, the observed luminosity-dependent peak redshift can emerge naturally from the joint DE+LE evolution acting on an LF whose slope varies with luminosity, without imposing \emph{a priori} different evolutionary laws for low- and high-power sources. This makes LADE particularly well suited for interpreting the intrinsic evolution of steep-spectrum radio-loud AGNs \citep[e.g.,][]{yuan2017mixture}. More broadly, related studies based on GMRT, LOFAR, and VLA samples have also explored the cosmic evolution of radio-AGN luminosity functions using a range of parametric approaches \citep[e.g.,][]{Ocran2021,2019A&A...622A..12H,2022MNRAS.513.3742K}. Notably, LADE has also been shown to provide a compelling description of the RLF of star-forming galaxies (SFGs) \citep{Novak_2017,2023MNRAS.523.6082C,2024A&A...683A.174W,2026ApJ...997..176W}, enabling like-for-like AGN--SFG comparisons within the same functional architecture.

In this work, we adopt the multi-survey parent sample compiled by \citet{2024A&A...684A..19S} and refine it into a beaming-minimized steep-spectrum subsample by excluding flat-spectrum sources. We then model the rest-frame 1.4\,GHz RLF within a LADE framework using a trivariate $(L,z,\alpha)$ RLF estimator that explicitly incorporates the spectral-index distribution.
Finally, we convert the steep-spectrum AGN RLF to a kinetic luminosity function to infer the radio-mode black hole accretion rate density (BHAD) traced by the steep-spectrum radio-loud AGN population, and compare its redshift dependence with the cosmic star formation rate density (SFRD) in a shape-based sense.

This paper is organized as follows. Section~\ref{sec_sample} describes the multi-survey datasets and the steep-spectrum sample construction, including the exclusion of flat-spectrum radio sources. Section~\ref{sec_methods} presents the binned and parametric methodology, completeness treatment, the LADE model families and model selection, and the joint constraints from local RLFs and source counts. Section~\ref{sec_results} presents the best-fit RLFs and evolutionary trends, including the inferred $z_{\rm peak}(L)$. Section~\ref{sec_discussion} derives BHAD and compares it with SFRD, and Section~\ref{sec_conclusion} summarizes our conclusions. Throughout this work, we assume a flat $\Lambda$ cold dark matter ($\Lambda$CDM) cosmology with $H_0 = 70\,\mathrm{km\,s^{-1}\,Mpc^{-1}}$, $\Omega_{\Lambda} = 0.7$, and $\Omega_{m} = 0.3$. We adopt a power-law radio spectrum $F_\nu \propto \nu^{-\alpha}$, with the spectral index fixed to $\alpha = 0.7$ where required.

\section{Data and Sample Selection}
\label{sec_sample}

\begin{figure}
	\centering
	\includegraphics[width=0.8\textwidth]{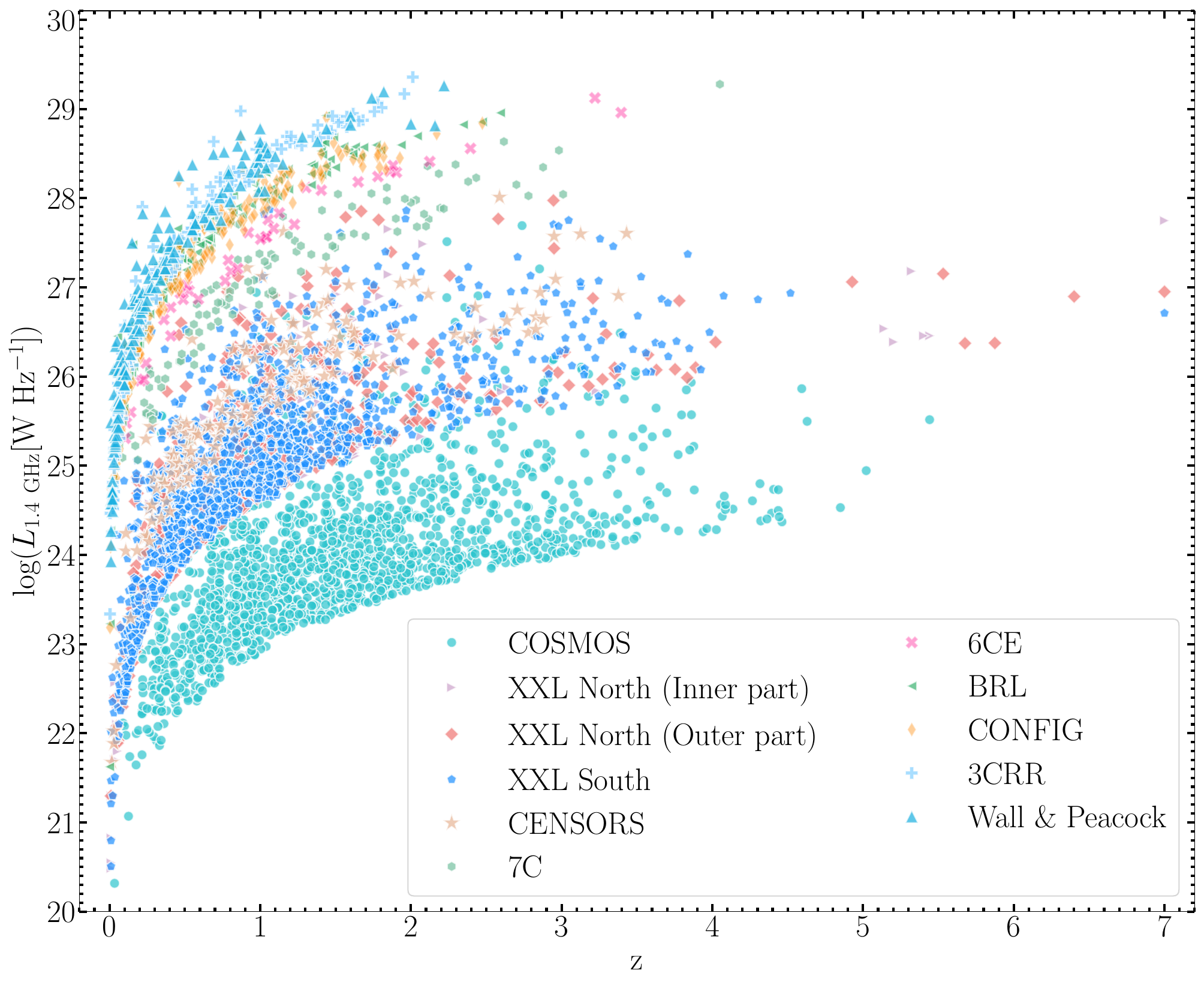}
    \caption{
    Redshift--luminosity distribution of the steep-spectrum radio-loud AGN sample used throughout this work (after excluding FSRQ sources). The corresponding survey fields are indicated in the legend.}
	\label{fig:sample_data}
\end{figure}

\begin{figure*}
    \centering
    \gridline{
        \fig{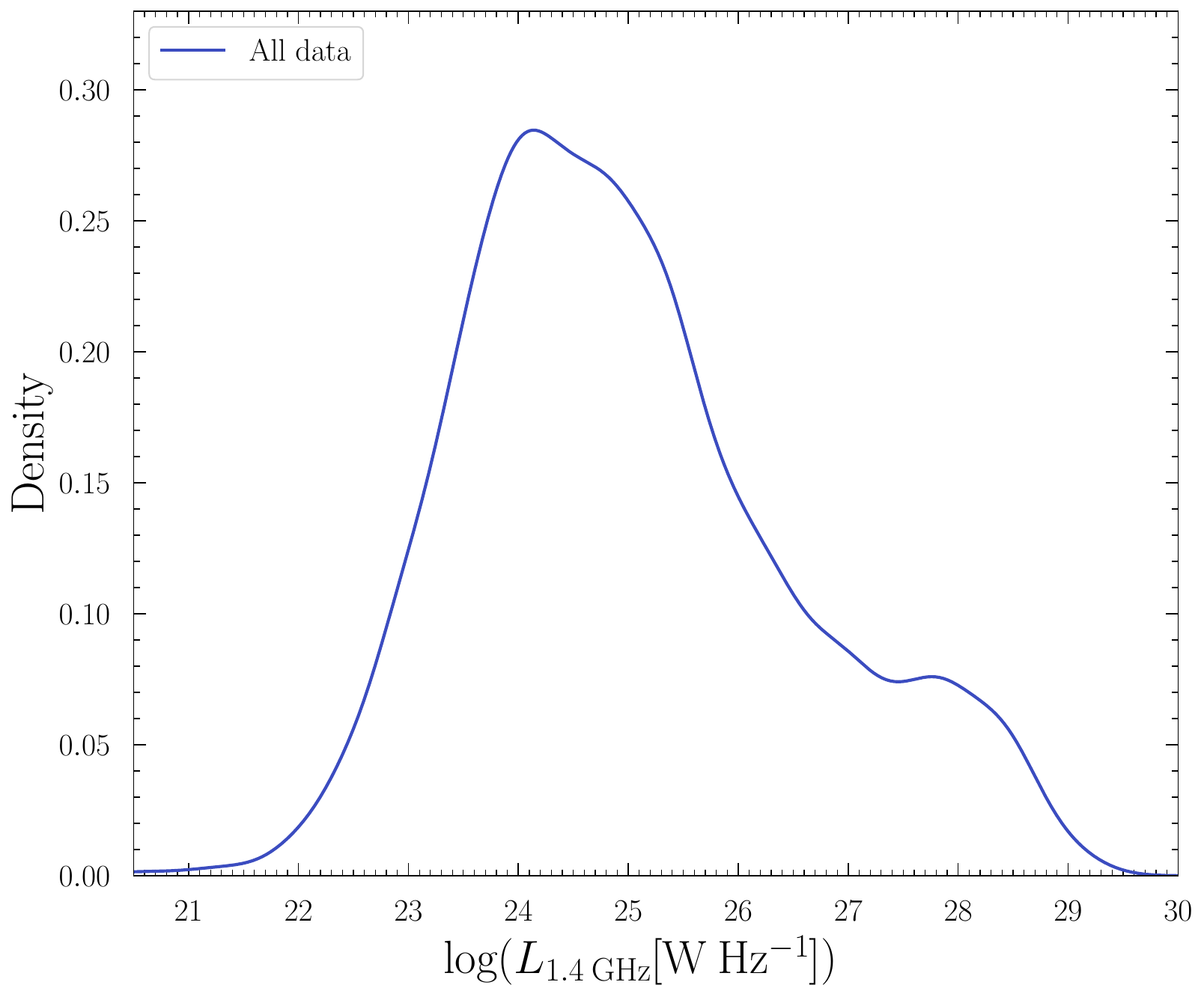}{0.5\textwidth}{(a)}
        \fig{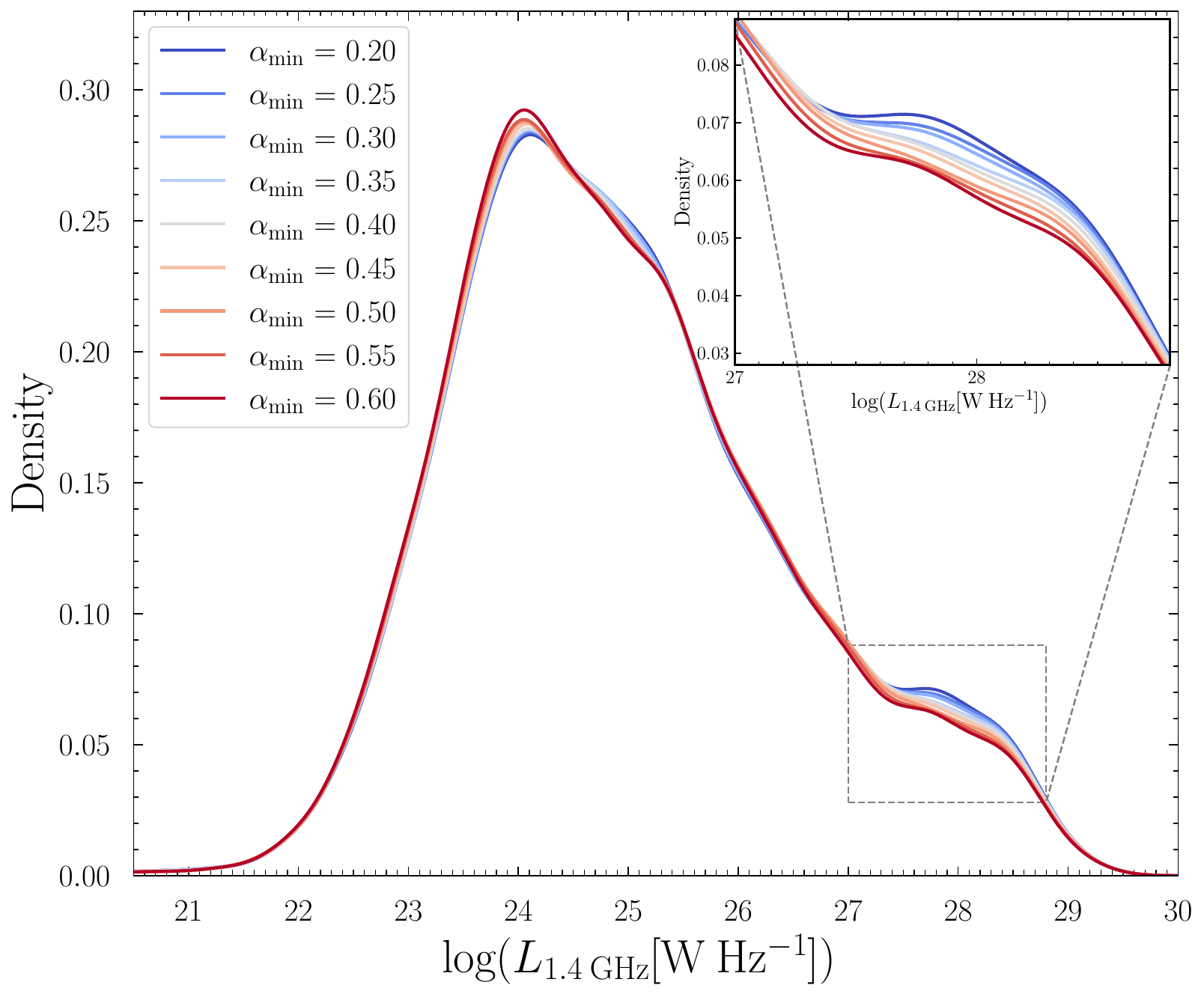}{0.5\textwidth}{(b)}
    }
    \vspace{-0.1in}
    \gridline{
        \fig{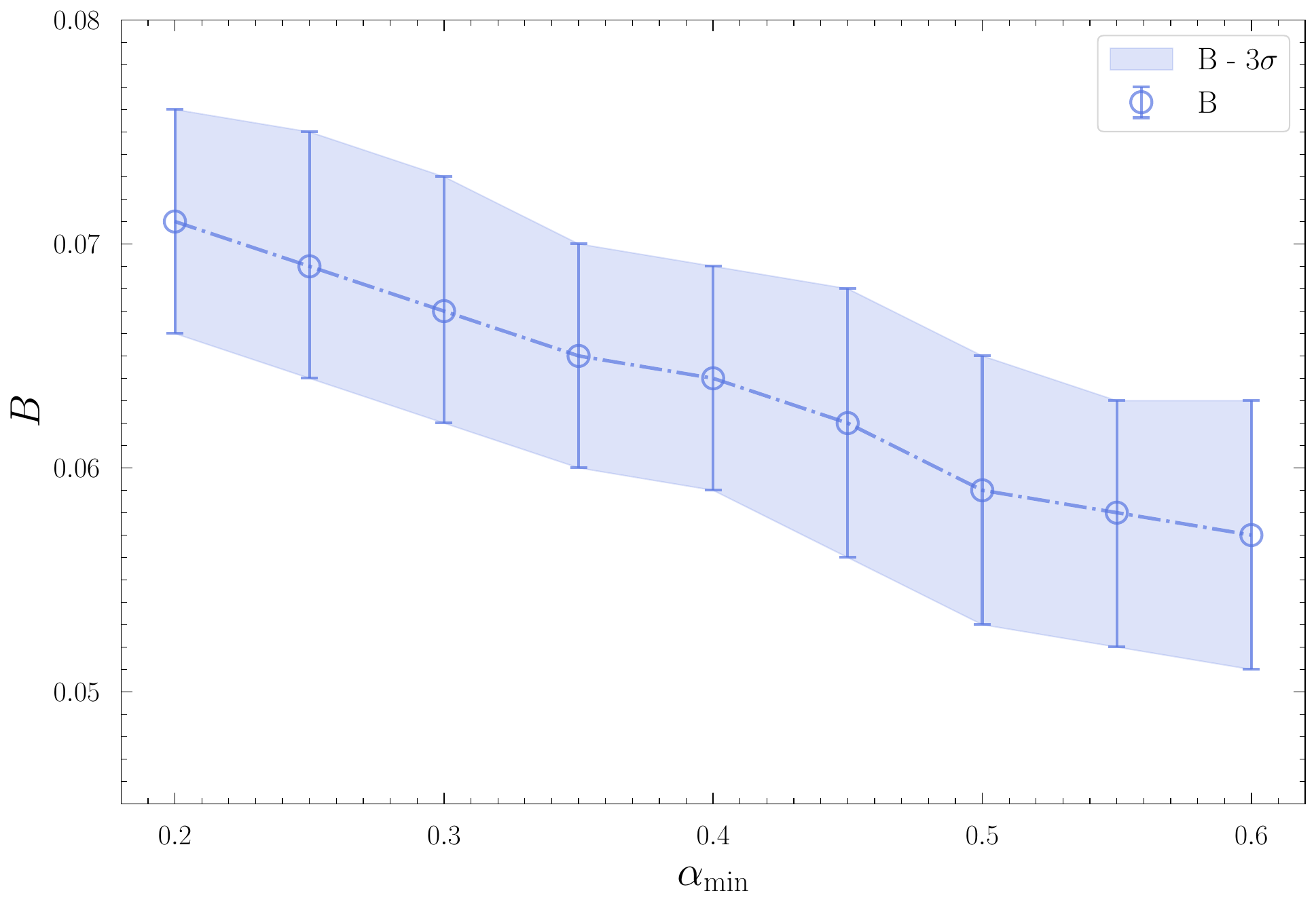}{0.5\textwidth}{(c)}
        \fig{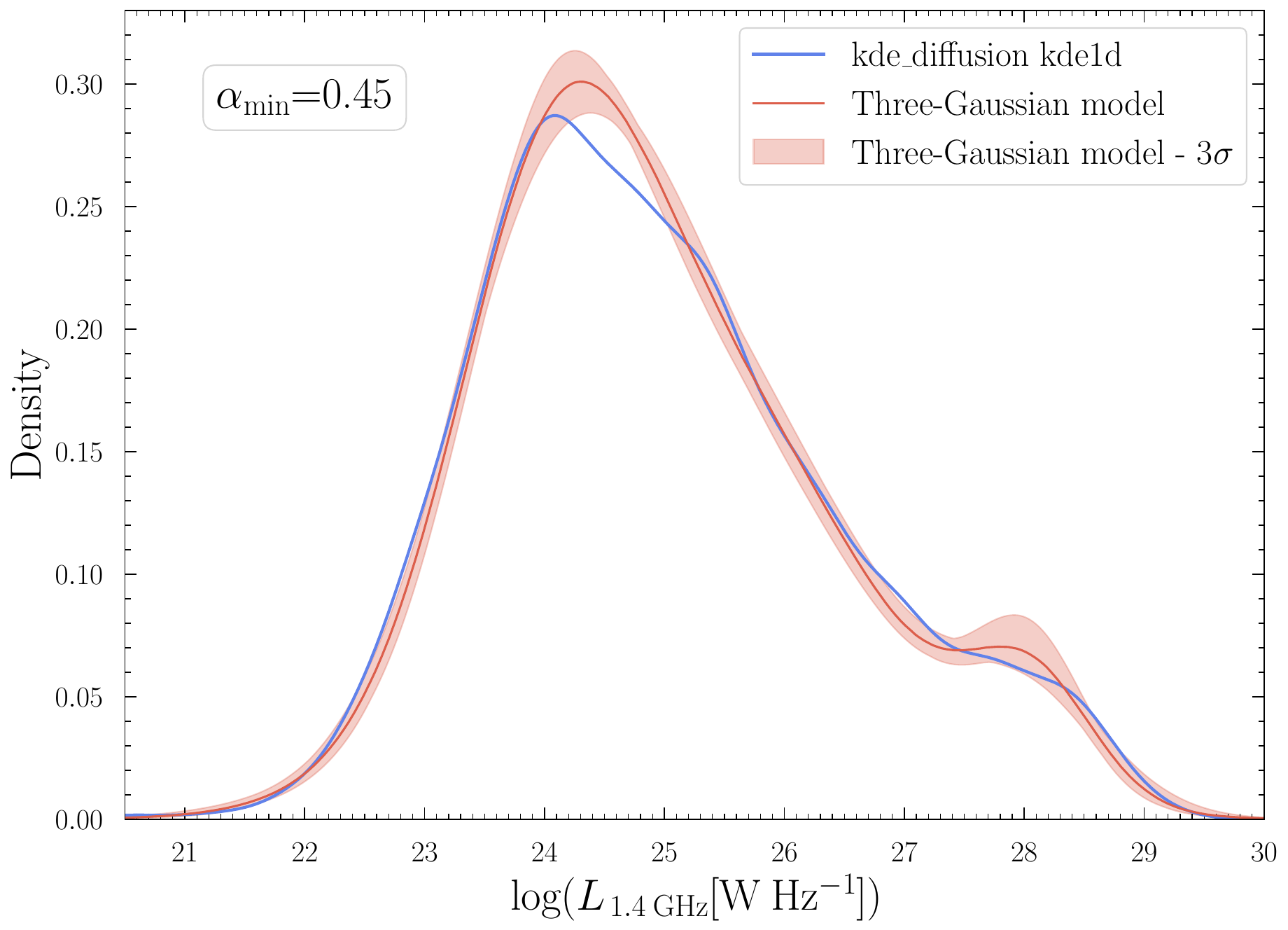}{0.5\textwidth}{(d)}
    }
    \caption{
    Panel (a): Kernel density estimate of the rest-frame 1.4\,GHz luminosity distribution
    for the full AGN sample, highlighting the presence of three distinct components.
    Panel (b): Evolution of the height of the high-luminosity peak as a function of the
    minimum spectral index threshold $\alpha_{\rm min}$, obtained by progressively removing
    sources with $\alpha < \alpha_{\rm min}$.
    Panel (c): Best-fitting weight $B$ of the FSRQ component in the Three-Gaussian model
    (Equation~\ref{eq:fsrq1}) as a function of $\alpha_{\rm min}$, with $3\sigma$ uncertainties.
    Panel (d): Comparison between the KDE result from \texttt{KDE-diffusion} and our
    Three-Gaussian model at $\alpha_{\rm min}=0.45$; the two distributions agree within
    the uncertainties.
    }
    \label{fig:sample_data2}
\end{figure*}

\begin{table*}
\centering
\caption{AGN data and FSRQ statistics across different surveys.}
\hspace*{-1.1cm}
\begin{tabular}{lcccc}
\hline\hline
Survey & Original Number & FSRQ Number & Final Number & FSRQ fraction (\%) \\
\hline
7C                    & 128  & 27  & 101  & 21.09 \\
6CE                   & 58   & 21  & 37   & 36.21 \\
3CRR                  & 170  & 26  & 144  & 15.29 \\
XXL-North (Inner)     & 292  & 86  & 206  & 29.45 \\
XXL-North (Outer)     & 607  & 150 & 457  & 24.71 \\
XXL-South             & 1484 & 279 & 1205 & 18.80 \\
COSMOS                & 1916 & 170 & 1746 & 8.87  \\
CENSORS               & 149  & 0   & 149  & 0.00  \\
BRL                   & 178  & 11  & 167  & 6.18  \\
Wall \& Peacock       & 227  & 64  & 163  & 28.19 \\
Config                & 243  & 63  & 180  & 25.93 \\
\hline
\hline
\end{tabular}
\label{tab:data}
\end{table*}

\subsection{Parent Sample}
The radio AGN sample utilized in this work is based on the comprehensive catalog compiled by \citet{2024A&A...684A..19S}, which originates from the XXL Survey Large Program. The data analysis and multi-band association procedure are fully described in that work, and we refer readers to that publication for a complete description. Here we summarize some key points about the sample.

This catalog incorporates data from a set of radio surveys with varying areas and depth.
The XXL Survey provides intermediate-depth coverage from both the XXL-North and XXL-South fields, observed with the Giant Metrewave Radio Telescope (GMRT) at 610 MHz and the Australia Telescope Compact Array (ATCA) at 2.1 GHz, respectively \citep{Smolcic2018, Butler2018}.
To capture both low- and high-luminosity AGN across a wide redshift range, the catalog includes additional wide and shallow surveys—3CRR, 6CE, and 7C—at 151 MHz \citep{Willott2001}, as well as deep-field observations from the VLA-COSMOS 3 GHz Large Project \citep{Smolcic2017b}. Further contributions come from the Combined EIS–NVSS Survey of Radio Sources (CENSORS) at 1.4 GHz \citep{Brookes2008}; the Best, R{\"o}ttgering, and Lehnert (BRL) sample at 408 MHz \citep{Best1999, 2000MNRAS.315...21B, Best2003}; the \citet{Wall1985} sample at 2.7 GHz; and the Combined NVSS–FIRST Galaxies (CoNFIG) sample at 1.4 GHz \citep{Gendre2008}, many of which include a high fraction of spectroscopic redshifts. To ensure consistency across surveys with different observing frequencies, all fluxes were converted to 1.4 GHz using a power-law spectrum and survey-specific spectral indices or their mean values \citep[see][Table 1]{2024A&A...684A..19S}.
In the parent catalogue of \citet{2024A&A...684A..19S}, the treatment of spectral indices is survey-dependent. For most surveys, spectral indices are taken from catalogue values derived from multi-frequency observations or from cross-matching with external radio catalogues (e.g., 7C, 6CE, 3CRR, XXL-N, XXL-S, COSMOS, Wall \& Peacock, and CoNFIG). For sources without an available source-level spectral index, the corresponding survey-mean value is adopted. The only survey explicitly assigned a fixed assumed value for all sources is CENSORS, for which $\alpha=0.8$ was used.

The AGN-selection steps described below are inherited from the parent catalogue of \citet{2024A&A...684A..19S}, rather than being newly performed in this work. In that parent catalogue, additional survey-dependent criteria were applied to construct a clean AGN sample. For the XXL-North and XXL-South fields, sources with flux densities above 1 mJy were retained to minimize contamination from SFGs. In the COSMOS field, AGN were selected based on a radio-excess criterion relative to infrared-based star formation rates. Bright-flux surveys (e.g., 3CRR, 6CE) required no additional cuts. Our own additional refinement is the subsequent exclusion of FSRQs, described in Section~\ref{sub_fsrq}, in order to construct the steep-spectrum subsample used for the RLF analysis.

The parent AGN sample before FSRQ exclusion consists of 5,452 AGNs spanning the redshift range $0 < z \lesssim 7$ and radio luminosities in the range $20 \lesssim \log(L_{\rm 1.4 GHz}) \lesssim 29.5$. A redshift--luminosity diagram of the sample is presented in Figure~\ref{fig:sample_data}, illustrating the ranges in redshift and luminosity that each survey spans. 
We also present in Figure \ref{fig:sample_data2}(a) the distribution of the rest-frame 1.4 GHz luminosity, $\log L_{\rm 1.4\,GHz}$, derived using the diffusion-based kernel density estimation (KDE) method of \citet{2010AnSta..38OS799B}, as implemented in the Python package \texttt{KDE-diffusion}\footnote{\url{https://pypi.org/project/KDE-diffusion/}}.
The AGN sample can be decomposed into three primary components: low-luminosity AGNs, high-luminosity AGNs, and flat-spectrum radio quasars (FSRQs). The latter population has been excluded from the analysis. A detailed explanation is provided below.

\subsection{Exclusion of Flat-Spectrum Radio Quasars}
\label{sub_fsrq}

Flat-spectrum sources are typically defined as those with a spectral index $\alpha < 0.5$, while steep-spectrum sources have $\alpha > 0.5$. Since the relativistic jets of FSRQs are oriented at small angles to our line of sight, Doppler boosting significantly enhances their observed luminosities relative to their intrinsic values. 
Consequently, their apparent luminosity distribution is strongly affected by orientation and selection effects, rather than reflecting the intrinsic population evolution of radio AGNs. For this reason, FSRQs are excluded from our analysis of the RLF. 

As shown in Figure~\ref{fig:sample_data2}(a), the luminosity distribution suggests a prominent tri-modal structure. Motivated by this empirical behavior, we describe it with a three-Gaussian mixture
\begin{eqnarray}
\label{eq:fsrq1}
\begin{aligned}
f(x) = &(1 - A - B) \cdot \frac{1}{\sqrt{2\pi}\sigma_1} \exp\left(-\frac{(x - \mu_1)^2}{2\sigma_1^2}\right) \\
&+ A \cdot \frac{1}{\sqrt{2\pi}\sigma_2} \exp\left(-\frac{(x - \mu_2)^2}{2\sigma_2^2}\right) \\
&+ B \cdot \frac{1}{\sqrt{2\pi}\sigma_3} \exp\left(-\frac{(x - \mu_3)^2}{2\sigma_3^2}\right),
\end{aligned}
\end{eqnarray}
where $x \equiv \log L_{\rm 1.4\,GHz}$. The parameters of this mixture are inferred by Markov Chain Monte Carlo (MCMC) sampling of the posterior probability distribution, assuming uniform priors and a likelihood defined for the one-dimensional luminosity-distribution data. We apply this three-Gaussian model to the full AGN sample as a practical preprocessing tool for identifying the high-luminosity FSRQ-related component, rather than as part of the final trivariate LF likelihood model used later in Section~\ref{sec_methods}.

From the MCMC fit, we obtain the best-fit peak positions of the three components as $\mu_1 = 24.04$, $\mu_2 = 25.139$, and $\mu_3 = 28.053$. In the subsequent analysis, these peak locations are held fixed when tracking the relative weight of the high-luminosity component as a function of the spectral-index threshold, under the working assumption that the characteristic locations of the three components remain approximately invariant. This preprocessing step is therefore not included in the parameter counting or model comparison of the final LF analysis.



As shown in Figure~\ref{fig:sample_data2}(b), by gradually increasing the lower limit on the spectral index $\alpha_{\rm min}$ (ranging from 0.2 to 0.6) and removing sources with $\alpha < \alpha_{\rm min}$, the height of the third peak decreases systematically. This indicates that the number of FSRQs declines with increasing spectral index threshold. 

We then applied the Three-Gaussian model described in Equation~\ref{eq:fsrq1} at each value of $\alpha_{\rm min}$ and performed MCMC fitting to obtain the corresponding weight $B$ of the FSRQ component together with its associated $3\sigma$ uncertainty. As illustrated in Figure~\ref{fig:sample_data2}(c), this weight also decreases with increasing $\alpha_{\rm min}$, exhibiting a consistent trend.

Based on these results, we adopt $\alpha_{\rm min} = 0.45$ as the final selection criterion to identify FSRQs. The choice of $\alpha_{\rm min}=0.45$ represents a balance between completeness and purity. A higher threshold would remove more FSRQs but might also exclude some genuine steep-spectrum AGNs due to spectral-index uncertainties, variability/non-simultaneous multi-frequency measurements, and spectral curvature, while a lower threshold would retain more sources but may fail to fully eliminate the FSRQ contamination. We note that flat-spectrum sources are not necessarily all strongly Doppler-boosted on an object-by-object basis, but they are statistically more susceptible to beaming-related luminosity inflation, motivating their removal to mitigate bias in the LF analysis. After excluding the FSRQ component, the final sample consists of 4,555 sources. The FSRQ statistics across the individual surveys are summarized in Table \ref{tab:data}. This refined dataset constitutes the steep-spectrum AGN sample utilized for the luminosity function analysis throughout the rest of this paper. Figure~\ref{fig:sample_data2}(d) compares the KDE result obtained with the diffusion-based estimator of \citet{2010AnSta..38OS799B} (via the \texttt{KDE-diffusion} package) with our Three-Gaussian model at $\alpha_{\rm min} = 0.45$. The two distributions agree within the uncertainties. 
We emphasize that this Three-Gaussian mixture is used only as a preprocessing tool for identifying and removing the FSRQ component when defining the steep-spectrum sample. It is not part of the final trivariate LF likelihood model fitted in Section~\ref{sec_methods}.

\section{Methods}
\label{sec_methods}
In this work, the AGN LF is estimated using both binned and parametric methods, which are briefly summarized below.

\subsection{The binned estimator}
\label{methods_pc}

To estimate the binned LF of AGNs, we adopt the method proposed by \citet{Page2000} (hereafter referred to as the PC method), which can be viewed as a practical binned variant of the classical $1/V_{\max}$ approach \citep{1968ApJ...151..393S}. Unlike the $1/V_{\max}$ method, which assigns each object an individual weight based on its maximum accessible volume, the PC method computes the LF by integrating over the \textit{actual surveyed region} of each luminosity--redshift ($L$--$z$) bin. 


For a given luminosity and redshift bin defined by $[L_1, L_2]$ and $[z_1, z_2]$, the LF at the center of the bin, as estimated by the PC method, is given by
\begin{equation}
\phi_{\mathrm{PC}} = \frac{N}{\displaystyle\int_{L_1}^{L_2} \int_{z_1}^{z_{\mathrm{max}}(L)} \frac{dV}{dz} \, dz \, dL},
\end{equation}
where $N$ is the number of AGNs observed in the bin, and the denominator reflects the volume actually accessible within the bin, constrained by the survey flux limit.
The uncertainty of each binned PC estimate is derived from the Poisson confidence interval of the number of sources in the corresponding bin. Writing the PC estimate as $\phi_{\rm PC}=N/V$, where $V$ denotes the accessible bin volume (i.e. the denominator of the PC estimator), we first compute the frequentist Poisson confidence interval of $N$ and then propagate it through the same normalization to obtain the upper and lower bounds on $\phi_{\rm PC}$. The error bars shown in the figures are therefore asymmetric in $\log \phi$, reflecting the propagated confidence interval in logarithmic space.
In the present work, the PC estimator is used only to provide a binned, non-parametric reference visualization of the LF. Our scientific inference is based on the unbinned trivariate likelihood analysis described in Section~\ref{sec:parametric_method}, rather than on the binned PC estimates themselves.

\subsection{The parametric method}
\label{sec:parametric_method}

The estimation of the AGN RLF can be significantly affected by the spectral index distribution of the sources \citep[see][]{2000MNRAS.319..121J,2001MNRAS.327..907J,2016ApJ...829...95Y}. As pointed out by \citet{2016ApJ...829...95Y}, neglecting this effect may lead to biased results, especially for flat-spectrum sources whose spectral indices exhibit large scatter. They demonstrated through Monte Carlo simulations that traditional bivariate estimators often fail to properly account for the complex truncation boundaries on the $L$--$z$ plane introduced by $K$-corrections. To address this issue, \citet{2016ApJ...829...95Y} proposed incorporating the spectral index distribution directly into the RLF analysis by introducing a trivariate function $\Phi(\alpha, z, L)$, which provides a more accurate and robust framework for estimating the RLF of AGNs. Motivated by their work, we adopt a similar parametric approach that explicitly takes into account the spectral index distribution when modeling the AGN RLF.

\subsubsection{The trivariate RLF}
\label{methods_LFLF}
Following \citet{yuan2017mixture}, we define the evolution of the trivariate RLF as the number density of sources per unit comoving volume $dV$ per unit logarithmic luminosity interval \( d\log L \), and per unit spectral index interval $d\alpha$
\begin{eqnarray}
\label{eq:lflf1}
\begin{aligned}
\Phi(\alpha,z,L)=\frac{d^{3}N}{d\alpha dV d\log L}.
\end{aligned}
\end{eqnarray}
If the spectral index distribution, $dN/d\alpha$, is independent of redshift and luminosity, the trivariate RLF can be factorized as the product of the conventional (bivariate) RLF, $\phi(z, L)$, and the intrinsic spectral index distribution:
\begin{eqnarray}
\label{eq:lflf2_revised}
\Phi(\alpha, z, L) = \phi(z, L) \times \frac{dN}{d\alpha}.
\end{eqnarray}
We adopt this factorization as a first-order approximation. We note that some observational studies have reported an apparent steepening of radio spectra with redshift (the so-called $z$--$\alpha$ relation), which has often been used to identify high-redshift radio sources. However, the strength of this trend, and even its physical origin, remain uncertain and are known to depend on sample-selection effects, observing frequency, and $K$-corrections \citep[e.g.,][]{2012MNRAS.420.2644K,2018MNRAS.480.2726M}. In the present work, our primary goal is to model the global evolution of the steep-spectrum radio-loud AGN RLF, and the available published parent catalog does not provide sufficient information to robustly constrain a redshift-dependent spectral-index distribution. We therefore retain a redshift-independent $dN/d\alpha$ as a pragmatic population-level approximation, while noting that \citet{2024A&A...684A..19S} found their LF results to be robust against different spectral-index treatments.

Given an analytical form of the LF parameterized by $\boldsymbol{\theta}$, denoted as $\Phi(\alpha,z, L\,|\,\boldsymbol{\theta})$, the optimal parameter set $\boldsymbol{\theta}$ is obtained by minimizing the negative log-likelihood function $S$. Following \citet{yuan2017mixture}, and adopting the general treatment of completeness corrections in \citet{marshall1983analysis} and \citet{fan2001high}, the likelihood function $S$ for a single survey field can be written in the unified form
\begin{eqnarray}
\label{eq:likelihood1}
\begin{aligned}
S_{\rm single} = 
&-2 \sum_{i=1}^{n} \ln \left[ \Phi(\alpha_{i}, z_{i}, L_{i}) \, p(z_{i}, L_{i}) \right] \\
&+ 2 \int\!\!\!\!\int\!\!\!\!\int_{W} \Phi(\alpha, z, L) \, p(z, L) \, \Omega \, \frac{dV}{dz} \, dz \, dL \, d\alpha ,
\end{aligned}
\end{eqnarray}
where $\Omega$ is the solid angle of the survey, $W$ denotes the survey region in the $(\alpha,z,L)$ space, and $p(z, L)$ is the completeness correction factor as a function of redshift and luminosity. In this formalism, a completely sampled field is treated as the special case with $p(z,L) \equiv 1$.

Among our eleven survey fields, data from eight fields (7C, 6CE, 3CRR, XXL-South, CENSORS, BRL, Wall \& Peacock, and Config) are considered to be complete, and we therefore set $p(z,L)\equiv 1$ for these fields. The remaining three fields—XXL-North (Inner), XXL-North (Outer), and COSMOS—are classified as incomplete. For the XXL-North field, the completeness factor $p(z,L)$ incorporates two principal components: one accounting for detection-limit noise \citep{Smolcic2018}, and another addressing redshift-dependent matching losses \citep{Slaus2020}. For the COSMOS field, the corresponding completeness correction is adopted from \citet{Smolcic2017a}, specifically their Figure~16 and Table~2. We compute $p(z,L)$ following the same procedure as described in \citet{2024A&A...683A.174W}.
Specifically, for each incomplete field we construct a two-dimensional grid in the $(L,z)$ plane (with a resolution of $50\times50$). At each grid point, the corresponding observed flux density at the survey frequency is computed from $(L,z)$ using the adopted spectral index. The relevant completeness components for that field are then evaluated by interpolation from the published completeness curves or tables, following the methods described in the corresponding survey papers. These components are combined to yield the value of $p(z,L)$ on the grid, and the completeness factor entering the likelihood integral at an arbitrary $(L,z)$ position is then obtained by two-dimensional interpolation over this precomputed grid.

This unified expression can be naturally generalized to the case of multiple survey fields indexed by $j$, each characterized by its own LF $\Phi_{j}$, completeness function $p_{j}(z,L)$, survey region $W_{j}$, and solid angle $\Omega_{j}$. The total likelihood function $S_{\rm all}$ that incorporates all fields is then
\begin{eqnarray}
\label{eq:likelihood2}
\begin{aligned}
S_{\rm all} =
&-2 \sum_{j} \sum_{i=1}^{n_{j}} 
   \ln \left[ \Phi_{j}(\alpha_{i}, z_{i}, L_{i}) \, p_{j}(z_{i}, L_{i}) \right] \\
&+ 2 \sum_{j} \int\!\!\!\!\int\!\!\!\!\int_{W_{j}} 
   \Phi_{j}(\alpha, z, L) \, p_{j}(z, L) \, \Omega_{j} \, \frac{dV}{dz} \, dz \, dL \, d\alpha ,
\end{aligned}
\end{eqnarray}
where $n_{j}$ is the number of sources in field $j$, and thus $n = \sum_{j} n_{j}$ is the total number of sources. In this notation, fields with complete data correspond to $p_{j}(z,L)\equiv 1$, while incomplete fields adopt the non-trivial completeness functions $p_{j}(z,L)$ described above.

To further constrain the model, we adopt the approach of \citet{Willott2001} and \citet{yuan2017mixture} by incorporating recent measurements of the local radio LF (LRLF) and SCs into the fitting procedure.  
Since both LRLF and SC are one-dimensional distributions, their influence on the fitting is quantified via the standard $\chi^2$ statistic
\begin{eqnarray}
\label{chi2}
\begin{aligned}
\chi^{2}=\sum_{i=1}^{n}\left(\frac{f_{\text {data } i}-f_{\bmod i}}{\sigma_{\text {data } i}}\right)^{2},
\end{aligned}
\end{eqnarray}
where $f_{\text{data}\,i}$ is the observed value in the $i$th bin, while $f_{\text{mod}\,i}$ and $\sigma_{\text{data}\,i}$ are the corresponding model prediction and observed error, respectively.  
Since $\chi^2$ is related to the likelihood via $\chi^2 = -2 \ln(\text{likelihood})$ \citep[i.e., the same form as $S$;][]{Willott2001}, we define a combined likelihood function $S_{\text{joint}}$ that incorporates constraints from all three components of data (AGN data, LRLF data, and SC data). The expression is
\begin{eqnarray}
\label{chi2all}
S_{\text{joint}}=S_{\text{all}} + \chi^2_{\text{LRLF}} + \chi^2_{\text{SC}},
\end{eqnarray}
where $\chi^2_{\text{LRLF}}$ and $\chi^2_{\mathrm{SC}}$ denote the value of $\chi^2$ for the local radio LFs and SCs, respectively.
Using Equation~(\ref{chi2all}), the best-fit parameters of the LF are obtained by numerically minimizing the objective function $S_{\text{joint}}$.  
Following the Bayesian framework adopted in our previous work \citep[e.g.,][]{2016ApJ...829...95Y}, we estimate both the optimal parameter values and their posterior distributions \citep[see also][]{yuan2017mixture}.  
We assume uniform (i.e., uninformative) priors for all parameters and perform MCMC sampling using the {\sc emcee} Python package \citep{foreman2013emcee}.

\subsection{Constraints from Local LFs and Source Counts}
\label{method_LLF_SC}

\begin{figure}
\centering
\includegraphics[width=0.8\textwidth]{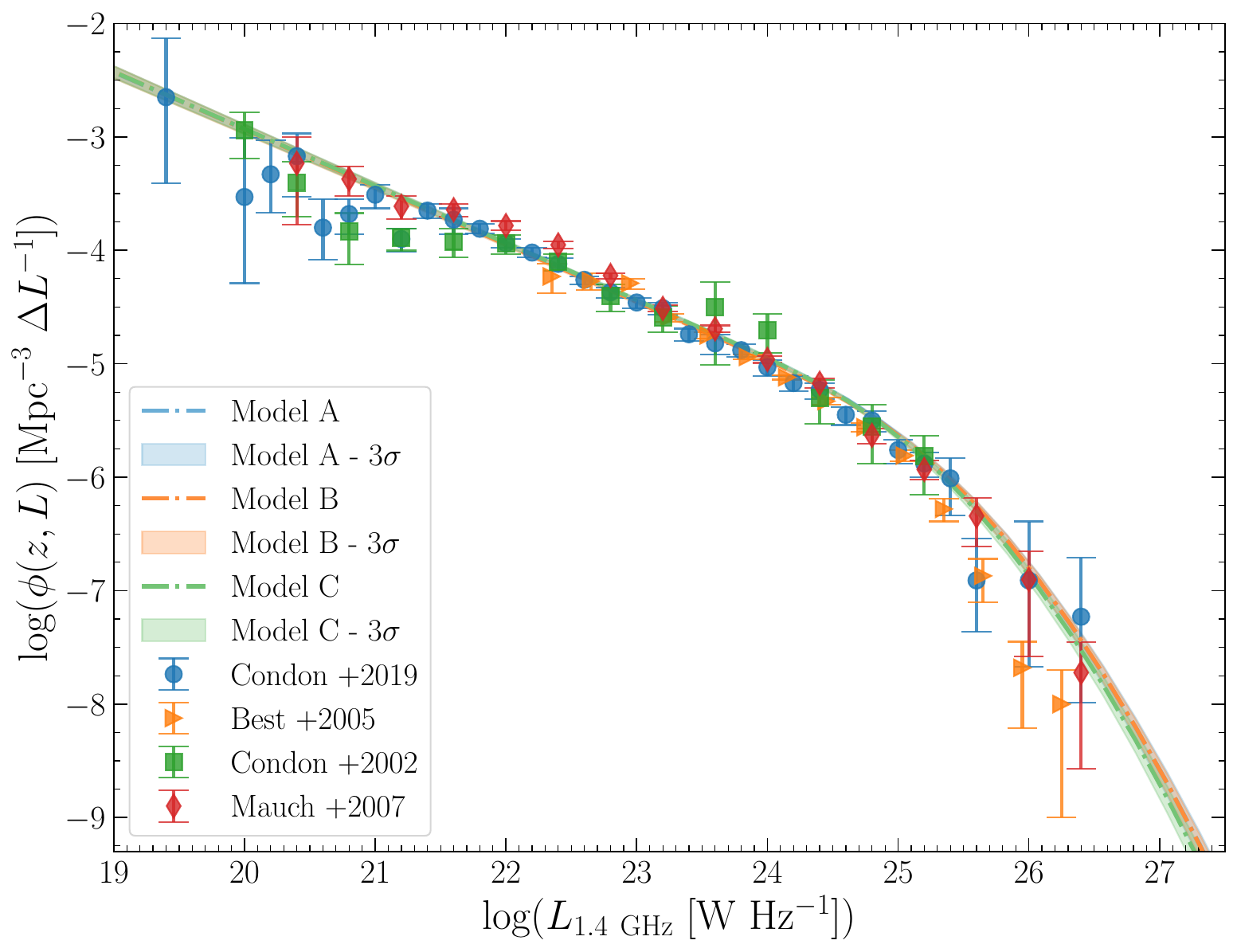}
\caption{
Local radio LF at 1.4 GHz of AGN from several surveys with different observed areas and sensitivities (colored data points). The colored curves show the $z=0$ evaluations of the globally fitted LADE models introduced in Section~\ref{method_LF}, using the corresponding best-fit parameter sets from the joint fit.}
\label{fig:LocalLuminosityFunction_LF}
\end{figure}

The local LFs at 1.4 GHz have been well determined for AGNs through the combination of large radio surveys, such as NVSS (NRAO VLA Sky Survey) and FIRST (Faint Images of the Radio Sky at Twenty centimeters), with wide-area
spectroscopic surveys. In this work, we jointly adopt the local AGN LFs reported by \citet{Condon_2002}, \citet{best2005sample}, \citet{Mauch_2007}, and \citet{Condon_2019} (see Figure \ref{fig:LocalLuminosityFunction_LF}) to calculate $\chi^2_{\text{LRLF}}$ in Equation (\ref{chi2all}), by summing the $\chi^2$ contributions from the individual literature local-LF datasets.

In addition to the local LFs, radio SCs provide important constraints for modeling the AGN LFs.  
The differential SCs, denoted as $n(F_\nu)$, describe the number of sources per unit flux density $F_\nu$ per steradian.  
Their distribution reflects both the cosmic evolution of the sources and the geometry of the Universe \citep{Padovani_2016}.  
The SCs are typically normalized to a Euclidean slope by multiplying by $F_\nu^{2.5}$ \citep[e.g.,][]{2010A&ARv..18....1D}.  
Following \citet{yuan2017mixture}, the AGN SCs can be related to their LF via
\begin{eqnarray}
\label{eq:sc1}
\begin{aligned}
\frac{n(F_\nu)}{4\pi}=4\pi\frac{c}{H_{0}}
\times\int_{z_{1}}^{z_{2}}\int_{\alpha_{1}}^{\alpha_{2}}\frac{\Phi(\alpha,z,L(\alpha,z,F_\nu))D^{4}_{L}(z)d\alpha dz}{(1+z)^{(3-\alpha)}\sqrt{\Omega_M(1+z)^3+\Omega_{\Lambda}}},
\end{aligned}
\end{eqnarray}
where $c$ is the speed of light, $\Phi(\alpha,z,L)$ is the LF, $D_L(z)$ is the luminosity distance, $z_1$ and $z_2$, $\alpha_1$ and $\alpha_2$ represent the integration limits of redshift and spectral index \citep{Padovani_2016}.

To further constrain our RLF model, we incorporate observed AGN radio SCs from 
\citet{Padovani2015}, \citet{Smolcic2017a}, \citet{Algera_2020}, and \citet{2023MNRAS.520.2668H}. 
We also include the bright-end radio source-count compilation of 
\citet{2010A&ARv..18....1D}. Because that compilation does not explicitly 
separate AGN and SFG contributions, we only use its data points at 
$F_\nu>1\,{\rm mJy}$, where the radio counts are generally expected to be AGN-dominated.
Taken together, these AGN-classified source counts and 
AGN-dominated bright-end counts span a wide range in frequency, flux density, 
and survey area, thereby providing complementary constraints on both the faint 
and bright ends of the AGN population. For consistency across all samples, SCs 
measured at frequencies other than 1.4\,GHz are converted to their equivalent 
1.4\,GHz values assuming a power-law radio spectrum.

These datasets are chosen for their complementary strengths. The compilation by \citet{2010A&ARv..18....1D} provides particularly broad coverage in flux density and frequency. From the E-CDFS VLA survey, \citet{Padovani2015} present a detailed decomposition of the sub-mJy radio population into radio-quiet (RQ) and radio-loud (RL) AGN based on extensive multiwavelength information, which we use to anchor the AGN contribution at intermediate flux densities. The VLA-COSMOS 3\,GHz Large Project \citep{Smolcic2017a} covers 2\,deg$^2$ and reaches a $5\sigma$ detection threshold of $\sim 11.5~\mu$Jy; we adopt the AGN SCs derived from their multiwavelength classification scheme to further constrain the faint end. In addition, we utilize the AGN SCs from the ultra-deep Karl G. Jansky VLA COSMOS-XS survey of \citet{Algera_2020}, which reaches an rms sensitivity of $\sim 0.53~\mu$Jy\,beam$^{-1}$ at 3\,GHz over an area of $\sim 350$\,arcmin$^2$. In that work, AGN are identified via multiwavelength diagnostics and subdivided into moderate-luminosity AGN (MLAGN) and high-luminosity AGN (HLAGN) according to their dominant emission mechanisms. Finally, \citet{2023MNRAS.520.2668H} present AGN SCs based on the 1.4\,GHz continuum Early Science data release of the MeerKAT MIGHTEE survey, which covers $\sim 5$\,deg$^2$ across the COSMOS and XMM-LSS fields; AGN are identified using multiwavelength diagnostics, and the SCs are corrected for incompleteness via extensive simulations.

A critical aspect of using the MIGHTEE counts from \citet{2023MNRAS.520.2668H} is the treatment of sources without secure classifications (``unmatched'' sources). They consider two schemes: (1) a flux-dependent probabilistic split between SFGs and AGN, based on the fractions observed in the classified sample within each flux bin, and (2) the extreme assumption that all unmatched sources are SFGs. In this work, we adopt the first, physically motivated approach, which assumes that the relative AGN fraction among unmatched sources mirrors that of the classified population in each flux bin. Accordingly, for both the COSMOS and XMM-LSS fields, we use the AGN SCs tabulated in the $\mathrm{SC}_{\mathrm{AGN,\,ratio}}$ column of Tables~1 and 2 in \citet{2023MNRAS.520.2668H}. We emphasize that the SC datasets incorporated into Equation~(\ref{chi2all}) are AGN-classified source counts rather than total radio-population counts. In particular, for Hale et al.~(2023) we use the $\mathrm{SC}_{\mathrm{AGN,\,ratio}}$ counts, as described above. Therefore, later comparisons to these same AGN SC datasets should be interpreted as consistency checks against the imposed constraints, rather than as independent validations.

\subsection{Models for the AGN LFs}
\label{method_LF}

Following \citet{yuan2017mixture}, the trivariate RLF can be defined as
\begin{eqnarray}
\label{aaa}
\Phi(\alpha,z,L)=e_1(z)\phi(z=0,L/e_2(z),\eta^j)\frac{dN}{d\alpha},
\end{eqnarray}
where $e_1(z)$ and $e_2(z)$ represent DE and LE with redshift, respectively, and $\eta^j$ denote the parameters that characterize the shape of the radio LF. If $\eta^j$ remains invariant across redshift, this suggests that the intrinsic shape of the radio LF is preserved. In contrast, a redshift-dependent $\eta^j$ implies luminosity-dependent density evolution \citep[for a more detailed discussion, see][]{singal2013radio,singal2014gamma}. In this work, we adopt the commonly used assumption that the LF shape does not vary with redshift (i.e., $\eta^j$ is fixed), consistent with several previous studies \citep[e.g.,][]{yuan2017mixture,Novak_2017,2022ApJ...941...10V,2024A&A...683A.174W}.

In this work, the local AGN LF $\phi(z=0, L/e_2(z=0))$ is modeled using the modified Schechter function proposed by \citet{saunders199060}:
\begin{eqnarray}
\label{eq:LF2}
\begin{aligned}
\phi(z&=0,L/e_2(z=0)) = \frac{dN}{d\log L} 
= \phi_{\star} \left( \frac{L}{L_{\star}} \right)^{1-\beta} 
\exp \left[ -\frac{1}{2\gamma^2} \log^2 \left(1 + \frac{L}{L_{\star}} \right) \right],
\end{aligned}
\end{eqnarray}   
where $L_{\star}$ marks the knee of the LF, $\beta$ and $\gamma$ control the slopes at the faint and bright ends, respectively, and $\phi_{\star}$ is the normalization factor of $\phi(z, L)$. 

This functional form is commonly used to describe the local LF of SFGs \citep[e.g.,][]{Novak_2017, Ocran2020, enia2022new, 2022ApJ...941...10V, 2023MNRAS.523.6082C, 2024A&A...683A.174W}. 
More recently, it has also been adopted to model the local LF of AGNs \citep[e.g.,][]{2024A&A...684A..19S}. In our analysis, we find that this form provides a better fit compared to other traditional prescriptions for AGN local LFs, such as the double power-law model \citep[e.g.,][]{2000MNRAS.317.1014B, 2012ApJ...751..108A, 2016ApJ...829...95Y} or the alternative modified Schechter function \citep[e.g.,][]{1995ApJ...438..623P, Hopkins2007, yuan2017mixture}.

In this work, we investigate three distinct evolutionary scenarios, designated as Model~A, Model~B, and Model~C. Models~A and B share an identical DE function, $e_1(z)$, parameterized as a simple exponential decay:
\begin{eqnarray}
\label{e1A}e_1(z)=e^{- p_1 z}.
\end{eqnarray}
The models are differentiated by their LE functions. For Model~A, we adopt a standard power-law evolution:
\begin{eqnarray}
\label{e2A}e_2(z) = (1 + z)^{k_1}.
\end{eqnarray}
In Model~B, to prevent the LE from increasing indefinitely with redshift, we introduce a redshift-dependent power-law index:
\begin{eqnarray}
\label{e2B}e_2(z) = (1 + z)^{k_1 + k_2 z}.
\end{eqnarray}
Finally, Model~C represents our most flexible scenario. It allows for a more complex density evolution structure than the simple monotonic decline assumed in Models A and B. Model~C retains the LE function of Model~B but adopts a DE function mirroring the mathematical form of the LE:
\begin{eqnarray}
\label{e1C}e_1(z) = (1 + z)^{p_1 + p_2 z}.
\end{eqnarray}
Notably, the functional forms of Model~B and Model~C are identical to those successfully applied to the SFG population in our previous work \citep{2024A&A...683A.174W}. These parameterizations fall under the general category of mixture evolution \citep[e.g.,][]{2016ApJ...829...95Y,yuan2017mixture} or LADE \citep[e.g.,][]{2010MNRAS.401.2531A} models.

Regarding the intrinsic spectral index distribution, $dN/d\alpha$, \citet{yuan2017mixture} demonstrated that a right-skewed distribution offers a superior description for steep-spectrum sources. Following their approach, we adopt the log-normal form:
\begin{eqnarray}
\label{SID}
\frac{dN}{d\alpha}=\frac{1}{\sqrt{2\pi} \alpha \sigma}e^{-\frac{(\ln\alpha-\mu)^2}{2\sigma^2}} .
\end{eqnarray}
Here, the source-level spectral-index values used in the sample construction and luminosity conversion are treated as fixed inputs, whether empirically measured or assigned from the corresponding survey mean when unavailable. The fitted $dN/d\alpha$ term in Equation~(\ref{SID}) therefore represents the intrinsic population-level spectral-index distribution, rather than a source-by-source free parameter in the likelihood optimization. 
Although the source-level spectral indices are fixed inputs, the observed one-dimensional distribution of these values is not necessarily an unbiased estimate of the intrinsic population-level \(dN/d\alpha\). In a flux-limited multi-survey sample, the probability that a source enters the sample depends on \(\alpha\) through the K-correction and through the conversion between the survey observing frequency and the rest-frame 1.4 GHz luminosity. Consequently, the accessible region in the \((\alpha,z,L)\) space is coupled to the luminosity--redshift selection boundary. Fitting \(dN/d\alpha\) independently from the observed \(\alpha_i\) distribution alone would therefore neglect the same truncation effects that motivate the trivariate LF formalism. For this reason, the parameters \(\mu\) and \(\sigma\) are constrained jointly with the LF and evolution parameters through the full likelihood, while the individual source-level \(\alpha_i\) values remain fixed observational inputs.
Accordingly, for each of Models A--C, the spectral-index-distribution parameters $(\mu,\sigma)$ are jointly constrained together with the four local-LF parameters $(\log\phi_\star,\log L_\star,\beta,\gamma)$ and the relevant evolution parameters through Equation~(\ref{chi2all}). The final trivariate LF models therefore have 8, 9, and 10 free parameters for Models A, B, and C, respectively. The assumption that the LF shape does not vary with redshift means that, within a given model, the shape parameters are redshift-invariant; it does not require the best-fit local-LF parameters to be identical across different evolutionary models. Since Models A, B, and C are fitted independently to the full dataset, they may yield different best-fit local-LF parameters even though each model assumes a redshift-invariant LF shape. The Three-Gaussian mixture introduced in Section~\ref{sub_fsrq} is a separate preprocessing step for FSRQ identification and is not included in the parameter counting of the final LF models.

\subsection{LDDE benchmark model}

For comparison with our LADE family, we also fitted a luminosity-dependent density evolution \citep[LDDE, e.g.,][]{1983ApJ...269..352S,2003ApJ...598..886U} benchmark model to the same steep-spectrum sample. Rather than directly adopting the published best-fit curve of \citet{2024A&A...684A..19S}, we adopt the same LDDE evolutionary prescription as in that work and re-fit it to our own FSRQ-excluded sample within the trivariate $(\alpha,z,L)$ framework described above. This ensures that the comparison between LADE and LDDE is carried out on the same dataset and under the same statistical framework.

Specifically, we write the trivariate LDDE luminosity function as
\begin{equation}
\Phi_{\rm LDDE}(\alpha,z,L)
=
\rho_{\rm LDDE}(z,L)\,
\phi(z=0,L)\,
\frac{dN}{d\alpha},
\end{equation}
where $\phi(z=0,L)$ is the local luminosity function given by Equation~(\ref{eq:LF2}), and $dN/d\alpha$ is the intrinsic spectral-index distribution given by Equation~(\ref{SID}). The luminosity- and redshift-dependent density evolution factor follows the same functional form as in \citet{2024A&A...684A..19S}:
\begin{equation}
\rho_{\rm LDDE}(z,L)
=
\frac{(1+z_c)^{q_1} + (1+z_c)^{q_2}}
{\left(\dfrac{1+z_c}{1+z}\right)^{q_1}
+
\left(\dfrac{1+z_c}{1+z}\right)^{q_2}},
\end{equation}
where
\begin{equation}
z_c=
\begin{cases}
z_c^\ast, & L>L_a\\[4pt]
z_c^\ast\left(\dfrac{L}{L_a}\right)^a, & L\leq L_a
\end{cases}
,
\end{equation}
where $z_c^\ast$ is the characteristic turnover redshift at high luminosities, $L_a$ is the transition luminosity, $a$ controls the luminosity dependence of the turnover redshift, and $q_1$ and $q_2$ describe the parameters of evolution.

In the LDDE benchmark, the four local-LF parameters $(\log\phi_\star,\log L_\star,\beta,\gamma)$, the five LDDE evolution parameters ($z_c^\ast$, $L_a$, $a$, $q_1$, $q_2$), and the two spectral-index-distribution parameters $(\mu,\sigma)$ are all treated as free parameters and are jointly constrained within the same likelihood framework; the LDDE benchmark therefore has 11 free parameters in total.
To ensure a like-for-like comparison with the LADE models, the LDDE benchmark is constrained using the same trivariate likelihood, completeness treatment, and joint fitting framework described in Sections~\ref{sec:parametric_method}--\ref{method_LLF_SC}.

\subsection{Model selection}
\label{method_aic}

To quantitatively compare the performance of different models, we adopt model selection criteria based on information theory \citep{takeuchi2000application}. One widely used metric is the Akaike Information Criterion \citep[AIC;][]{1974ITAC...19..716A}, which balances model fit and complexity. It is defined as
\begin{equation}
\label{eq:aic}
\text{AIC} = S_{\mathrm{all}}(\hat{\theta}) + 2q,
\end{equation}
where $S_{\mathrm{all}}(\hat{\theta})$ is the total likelihood statistic evaluated at the best-fit parameters $\hat{\theta}$, and $q$ denotes the number of free parameters.

An alternative criterion is the Bayesian Information Criterion \citep[BIC;][]{Schwarz1978}, which introduces a stronger penalty for model complexity:
\begin{equation}
\label{eq:bic}
\text{BIC} = S_{\mathrm{all}}(\hat{\theta}) + q \ln n,
\end{equation}
where $n$ is the effective sample size. For model comparison we use only the AGN LF likelihood, i.e. AIC and BIC are computed from $S_{\rm all}$ without including the $\chi^2$ constraints from the LRLF and source counts.
To guide the interpretation of the model-comparison statistics, we follow the commonly used empirical scale discussed by \citet{Kass1995}, according to which a difference of $\Delta\mathrm{BIC}>8$ is generally regarded as strong evidence against the model with the larger BIC. In the present work, the same qualitative ranking is also supported by the corresponding AIC differences.

\begin{table*}
\centering
\caption{Model comparison based on information criteria. The left panel shows the AIC and BIC statistics for the full sample, while the right panel presents the AIC statistics for the restricted subset ($0 < z < 3.5$) to test the robustness of the models against high-redshift selection effects. $\Delta$ values are computed relative to Model~C, which provides the minimum values in both cases. For reference, $\Delta\mathrm{BIC}>8$ is commonly interpreted as strong evidence against the model with the larger BIC \citep{Kass1995}.}
\begin{tabular}{l cccc c cc}
\hline\hline
 & \multicolumn{4}{c}{Full Sample} & & \multicolumn{2}{c}{Subset ($0 < z < 3.5$)} \\
\cline{2-5} \cline{7-8} \\[-2ex] 
Model~~ & AIC & $\Delta$AIC & BIC & $\Delta$BIC & & AIC & $\Delta$AIC \\
\hline
A & 126913.9 & 44.2 & 126965.3 & 31.3 & & 125255.3 & 49.1 \\
B & 126888.2 & 18.5 & 126946.0 & 12.0 & & 125228.6 & 22.4 \\
C & 126869.7 & ~~0.0 & 126934.0 & ~~0.0 & & 125206.2 & ~~0.0 \\
LDDE & 127580.5 & 710.8 & 127651.2 & 717.2 & & -- & -- \\
\hline
\end{tabular}
\label{table:aicbicpara}
\end{table*}

\begin{figure*}
\centering
\includegraphics[width=0.85\textwidth]{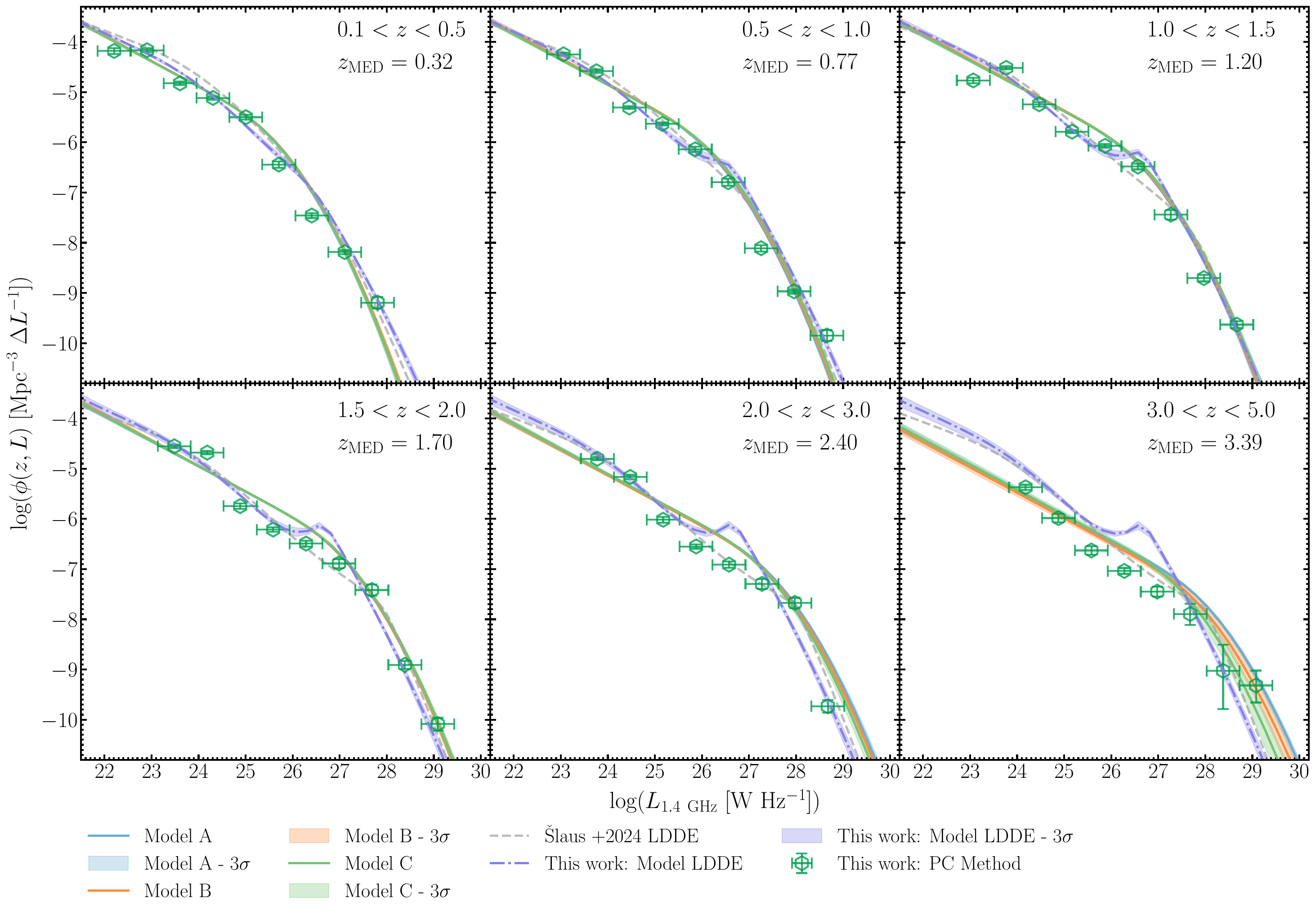}
\includegraphics[width=0.85\textwidth]{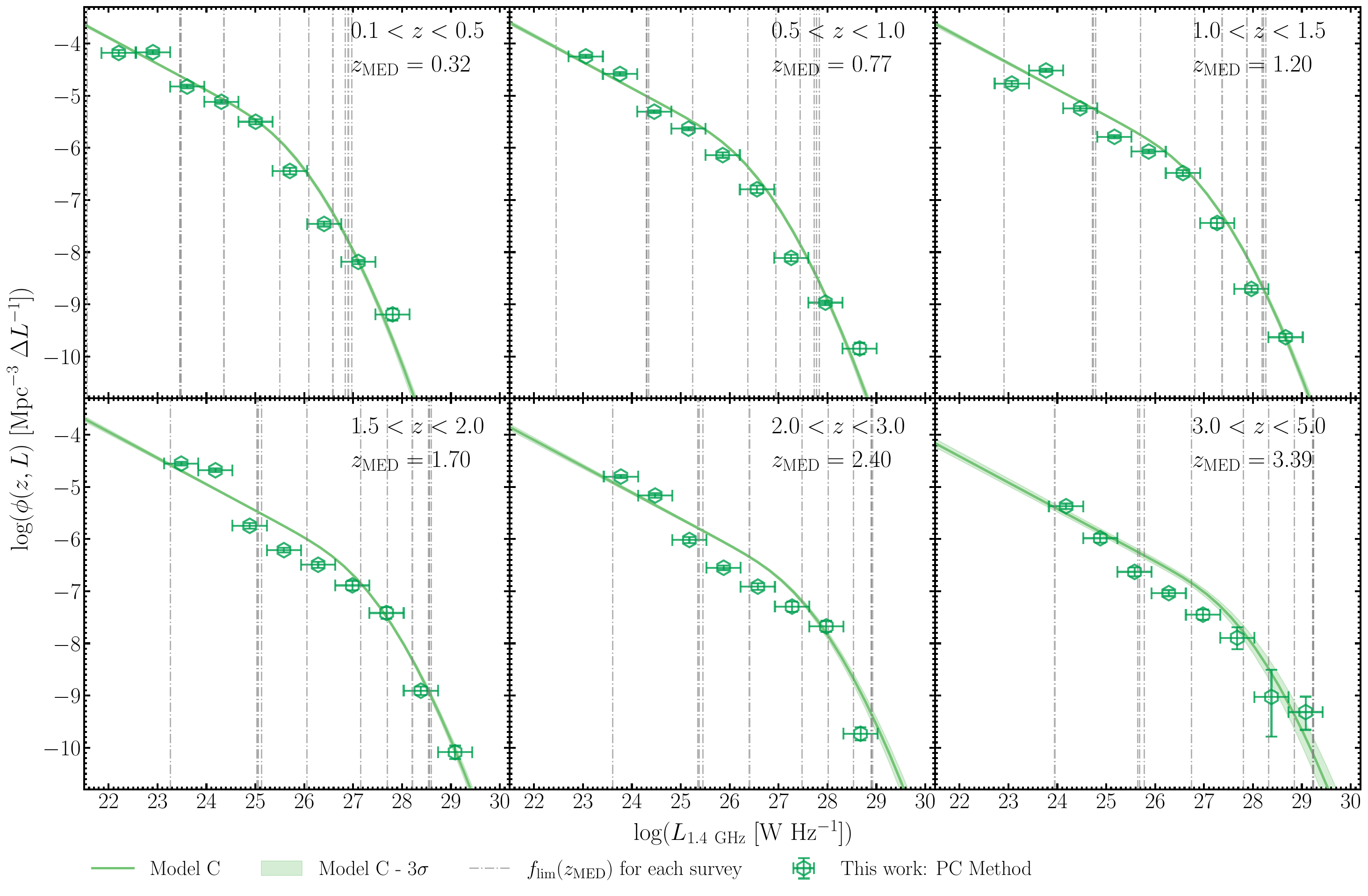}
\caption{
Radio LFs of AGNs in different redshift bins, compared with the nonparametric LFs obtained using the PC method (green open hexagons). The best-fitting parametric LFs for Models A, B, and C are shown as blue, orange, and green solid curves, respectively, with shaded bands indicating the corresponding $3\sigma$ confidence regions. For comparison, the grey dashed curves show the LDDE model of \citet{2024A&A...684A..19S}, while the purple dash-dotted curves depict our own trivariate LDDE fit obtained by re-fitting the same LDDE evolutionary prescription to the steep-spectrum sample used in this work}, with their $3\sigma$ confidence intervals indicated by the purple shaded regions. The top panels display the full comparison among Models~A, B, and C, whereas the bottom panels only show Model~C and its $3\sigma$ confidence region together with the PC-based LFs, to highlight the agreement between our fiducial model and the nonparametric estimates. The vertical grey dash-dotted lines in the bottom panels mark the flux limits of the individual surveys contributing to each redshift bin.

\label{fig:LF_PC}
\end{figure*}

\section{Results}
\label{sec_results}

In this section, we present the derived AGN radio LF, quantify their evolution across cosmic time, and compare the results from our LADE model with an alternative LDDE model. We also explore how our findings relate to existing work on the luminosity-dependent evolution of radio AGN and, finally, examine evidence for a coevolution between AGN and SFGs.
The fitting workflow is as follows. For each parametric model, we directly fit the trivariate RLF to the source-level data by minimizing the joint statistic $S_{\rm joint}$ in Equation~(\ref{chi2all}), which includes the AGN sample together with the external LRLF and source-count constraints. The model curves shown in Figures~3 and 4 are then obtained from the corresponding best-fit parameter sets. The binned PC estimates are computed independently and are used only as a non-parametric reference, not as the primary fitting target.

\begin{table*}
\tablewidth{0pt}
\renewcommand{\arraystretch}{1.5} 
\setlength{\tabcolsep}{1mm}       
\caption{Best-fit parameters for models A, B, and C}
\centering
\resizebox{\textwidth}{!}{%
\hspace*{-3.4cm} 
\begin{tabular}{lccccccccccc}
\hline\hline
Model~~ 
& $\log_{10}(\phi_\star)$ & $\log_{10}(L_{\star})$ 
& $\beta$ & $\gamma$ & $k_1$ & $k_2$ & $p_1$ & $p_2$ 
& $\mu$ & $\sigma$ \\
\hline
A~~ 
& $-5.043_{-0.023}^{+0.023}$ & $24.115_{-0.038}^{+0.038}$ 
& $1.503_{-0.004}^{+0.004}$ & $0.862_{-0.009}^{+0.009}$ 
& $4.661_{-0.072}^{+0.072}$ & \multicolumn{1}{c}{\ldots} & $1.330_{-0.024}^{+0.024}$ & \multicolumn{1}{c}{\ldots} 
& $-0.235_{-0.005}^{+0.005}$ & $0.280_{-0.004}^{+0.004}$ 
\\

B~~ 
& $-5.080_{-0.026}^{+0.026}$ & $24.147_{-0.041}^{+0.041}$ 
& $1.507_{-0.004}^{+0.004}$ & $0.853_{-0.0 10}^{+0.010}$
& $4.915_{-0.090}^{+0.090}$ & $-0.151_{-0.036}^{+0.036}$
& $1.296_{-0.027}^{+0.027}$ & \multicolumn{1}{c}{\ldots} 
& $-0.235_{-0.005}^{+0.005}$ & $0.282_{-0.004}^{+0.004}$ \\

C~~ 
& $-4.985_{-0.031}^{+0.031}$ & $24.024_{-0.049}^{+0.049}$ 
& $1.503_{-0.004}^{+0.004}$ & $0.848_{-0.011}^{+0.011}$
& $6.010_{-0.180}^{+0.180}$ & $-0.583_{-0.069}^{+0.069}$ 
& $-2.220_{-0.120}^{+0.120}$ & $-0.149_{-0.043}^{+0.043}$ 
& $-0.236_{-0.005}^{+0.005}$ & $0.282_{-0.004}^{+0.004}$ \\
\hline
\end{tabular}}
\vspace{0.2cm}
\parbox{\textwidth}{%
    \raggedright\small
    Note. Units --- $\phi_\star$: [${\rm Mpc^{-3}~dex^{-1}}$], 
    $L_{\star}$: [${\rm W~Hz^{-1}}$]. 
    The best-fit parameters and their $1\sigma$ uncertainties are shown 
    for models A, B, and C.
    The parameters listed here are obtained from independent global fits of Models A, B, and C to the full dataset via Equation~(\ref{chi2all}); they are therefore not expected to be identical across different evolutionary models.}%
\label{modelpara}
\end{table*}

\subsection{Best-fit Luminosity Functions}

The parameters of our model LFs are constrained using the MCMC algorithm implemented via the {\sc{emcee}} Python package \citep{foreman2013emcee}. This ensemble sampler employs multiple chains initialized from different starting points, allowing for an efficient and robust exploration of the parameter space while avoiding local minima. We adopt uniform priors for all model parameters. The one- and two-dimensional marginalized posterior distributions for Models A, B, and C are presented in Figures~\ref{fig:cornerplotA}--\ref{fig:cornerplotC}, demonstrating that all parameters are well constrained. The best-fit values along with their corresponding 1$\sigma$ uncertainties are summarized in Table~\ref{modelpara}.



Figure~\ref{fig:LF_PC} presents the best-fit RLFs derived from our Models A (blue), B (orange), and C (green solid lines), together with the binned estimates from the PC method (green open hexagons). In this plot, all RLFs are evaluated at the rest-frame frequency of 1.4 GHz. We also plot the original LDDE model from \citet{2024A&A...684A..19S} (grey dashed line). Since that model was derived from the broader parent sample containing FSRQs, it is not strictly comparable to our steep-spectrum results. To enable a rigorous comparison, we fitted the same LDDE functional form to our defined steep-spectrum sample, accounting for the intrinsic spectral index distribution. The resulting fit, referred to as our trivariate LDDE model (purple dash-dotted line), serves as a robust baseline for evaluating our LADE models.

Qualitatively, our three models are virtually indistinguishable across the entire luminosity range at $z \lesssim 2.5$. It is only at higher redshifts ($z\gtrsim2.5$) that minor differences begin to emerge at the high-luminosity end. Moreover, our LADE-based models show good overall agreement with the PC-derived binned RLFs, reproducing the main trends within the uncertainties; however, we note that the binned data points lie systematically below the model curves in several luminosity bins. This behavior is expected and can be attributed to two primary methodological effects discussed below.

First, while it is common practice to regard binned estimates as `observed LFs' and use them as a benchmark to validate parametric models, we must emphasize a critical caveat regarding our composite sample. Since our dataset combines 11 sub-surveys with distinct flux limits, the flux-limit curves of these different fields inevitably cross multiple luminosity bins within any given redshift interval. Consequently, the resulting binned LFs are highly sensitive to the choice of binning—a manifestation of boundary bias \citep[e.g.,][]{Yuan2013}. While such bias is typically confined to the faintest bins in a single-flux-limit sample, the multiple thresholds in our composite dataset mean that these boundary effects can manifest across a much broader luminosity range. In this work, to facilitate direct comparison, we have adopted the identical binning strategy as \citet{2024A&A...684A..19S}. As a consequence, the binned PC estimates may show localized bin-to-bin features (including apparent ``bumps'') that depend on the adopted binning and on the crossing of multiple flux-limit boundaries, whereas our parametric curves are intentionally smooth and are not expected to reproduce every such local fluctuation.

A second, related factor is that the PC estimator does not explicitly account for the intrinsic distribution of spectral indices. The visibility of a source depends not only on its luminosity and redshift but also on its spectral index via the K-correction. In bins intersected by the flux-limit curves, sources with `unfavorable' spectral indices (i.e., those resulting in lower observed fluxes) may fall below the detection threshold, even if they theoretically reside within the accessible (L,z) volume defined by a mean spectral index. While our trivariate parametric models fully integrate over this $\alpha$-distribution to predict the total space density, the PC estimator simply counts the detected sources without compensating for this spectral-dependent incompleteness. We note that the classical $1/V_{\text{max}}$ estimator \citep{1968ApJ...151..393S} is subject to the identical limitation. This inevitably leads to an underestimation of the luminosity function in these boundary-crossing bins \citep[see][]{2016ApJ...820...65Y}, further explaining why the PC data points frequently lie below our model curves. In individual boundary-crossing bins, these effects can also introduce additional bin-to-bin structure in the binned estimates.

Finally, we examine the behavior of our trivariate LDDE model (purple dash-dotted line). Note that the pronounced ``bump'' discussed below is a feature of the LDDE fit itself, and is distinct from any localized bin-to-bin structure that may appear in the binned PC points. At lower redshifts ($z\lesssim 1.5$), this model shows good agreement with our LADE predictions. However, as redshift increases, the complex effects of luminosity-dependent density evolution become increasingly prominent. Specifically, the LDDE model begins to predict space densities exceeding those of the LADE framework at the faint end ($\log L\lesssim 25$). Most notably, a significant `bump' emerges in the intermediate luminosity range ($26\lesssim\log L\lesssim 27$), which is immediately followed by a steeper decline at the bright end. This dramatic variation in the shape of the luminosity function suggests a scenario where low- and high-luminosity sources undergo markedly different evolutionary trajectories. However, the PC estimates do not appear to support such a scenario. 

This visual assessment is strongly corroborated by the quantitative model selection criteria. The resulting AIC and BIC values are summarized in the left panel of Table~\ref{table:aicbicpara}. We find that the LADE framework provides a statistically superior description of the data compared to the LDDE parameterization, which yields significantly higher information criterion values ($\Delta \mathrm{AIC} > 700$ relative to Model C). Among the three LADE variants, Model C consistently achieves the lowest AIC and BIC scores. This indicates that the LADE model with flexible density and luminosity evolution (Model C) offers the most parsimonious representation of the cosmic evolution of steep-spectrum radio-loud AGNs, optimally balancing goodness-of-fit with model complexity.

\subsection{Robustness of the Evolutionary Model}

A potential concern in modeling the high-redshift evolution is the impact of observational selection effects. Specifically, at $z \gtrsim 3.5$, our composite sample is dominated by deep, pencil-beam surveys (e.g., COSMOS), which may lack the volume to detect rare, high-luminosity sources. One might suspect that the statistical preference for Model C—which allows for a flexible turnover in density and luminosity evolution—is primarily driven by the need to accommodate this high-redshift truncation.

To test the robustness of our results, we performed a sensitivity analysis by repeating the MCMC fitting procedure on a restricted subset of data with $z < 3.5$, where the coverage in the luminosity-redshift plane is most complete. The resulting AIC values for this subset are summarized in the right panel of Table~\ref{table:aicbicpara}.

Remarkably, Model C remains the statistically preferred description, with a $\Delta \text{AIC}$ of $22.4$ relative to Model B. This confirms that the complex evolutionary trends captured by Model C (including the non-monotonic behavior in luminosity or the flattening in density evolution) are already firmly established by the high-quality data around `cosmic noon' ($z \sim 1-3$) and are not merely artifacts of the sparse sampling at the highest redshifts. 
Consequently, we conclude that Model C is the statistically preferred model, effectively characterizing the evolutionary behavior of radio-loud AGNs.

\begin{figure}
	\centering
	\hspace{-0.2in}
	\includegraphics[width=0.8\textwidth]{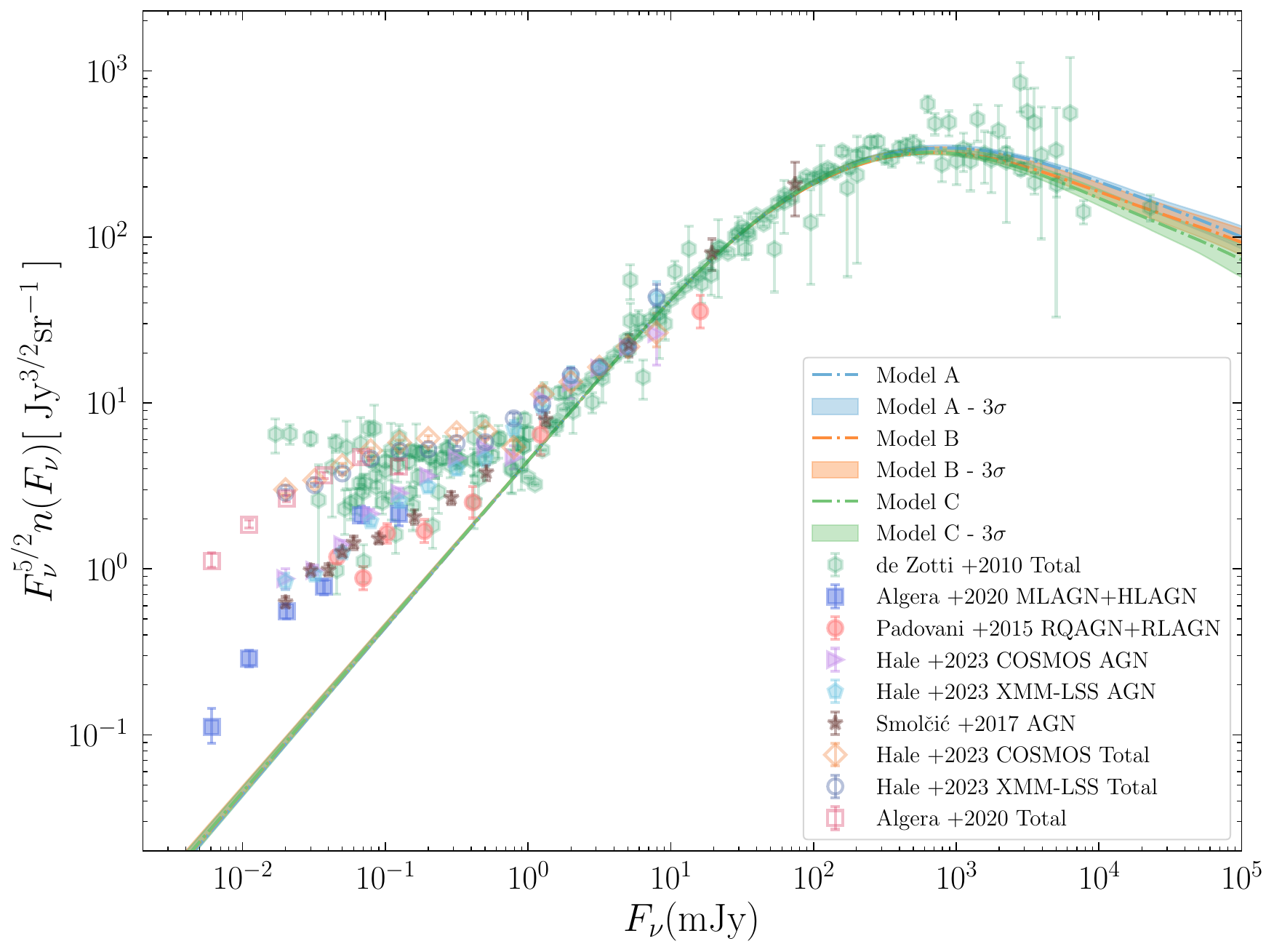}
	\caption{
    Euclidean-normalized differential source counts at 1.4~GHz predicted by the best-fitting LADE models and compared with observed radio source-count measurements.
    The blue, orange, and green dash-dotted curves show the predictions from Models A, B, and C, respectively, with the corresponding shaded regions indicating the $3\sigma$ confidence intervals.
    The AGN-classified source counts are shown as blue squares for the MLAGN+HLAGN counts of \citet{Algera_2020}, red circles for the RQAGN+RLAGN counts of \citet{Padovani2015}, purple right-pointing triangles for the COSMOS AGN counts of \citet{2023MNRAS.520.2668H}, cyan pentagons for the XMM-LSS AGN counts of \citet{2023MNRAS.520.2668H}, and brown five-pointed stars for the AGN counts of \citet{Smolcic2017a}.
    For reference, total radio source counts are also shown as light-green hexagons for the bright-end compilation of \citet{2010A&ARv..18....1D}, orange open diamonds for the COSMOS total counts of \citet{2023MNRAS.520.2668H}, pale-blue open circles for the XMM-LSS total counts of \citet{2023MNRAS.520.2668H}, and pink open squares for the total counts of \citet{Algera_2020}.
    }
	\label{fig:Source_Counts}
\end{figure}

\begin{figure*}
    \centering
    \gridline{
        \fig{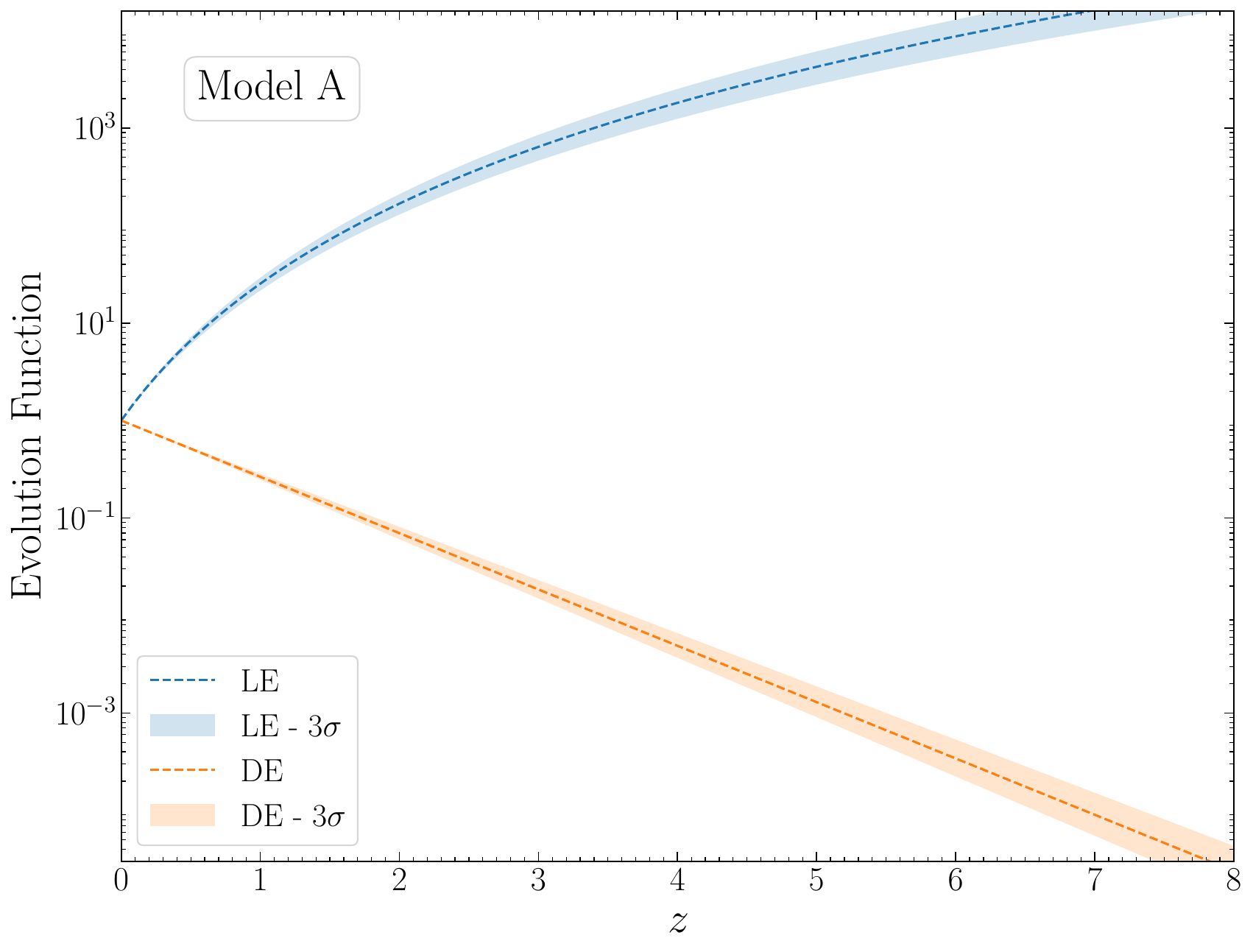}{0.33\textwidth}{}
        \hspace{-0.1in}
        \fig{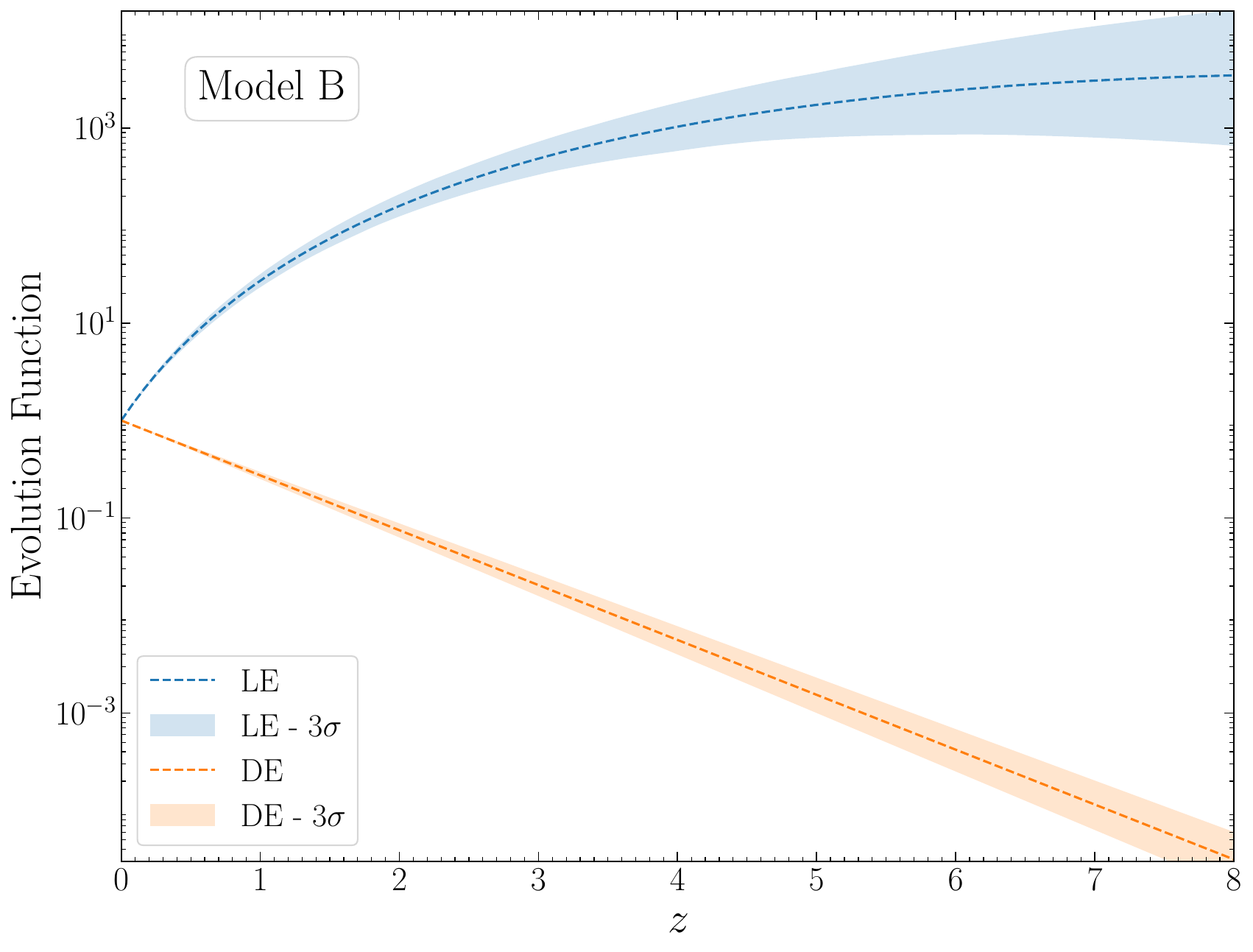}{0.33\textwidth}{}
        \hspace{-0.1in}
        \fig{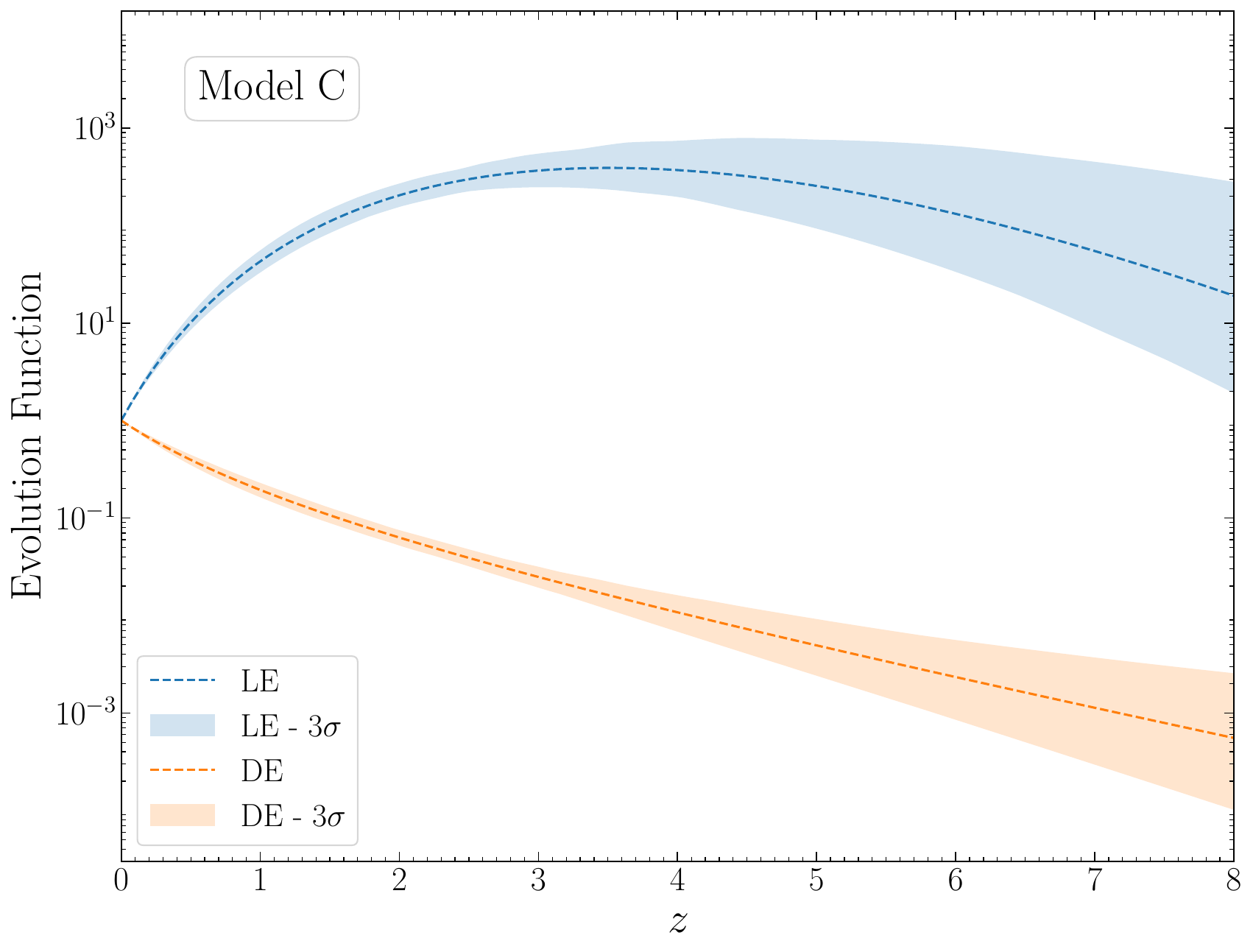}{0.33\textwidth}{}
    }
    \vspace{-0.4in} 
    \gridline{
        \fig{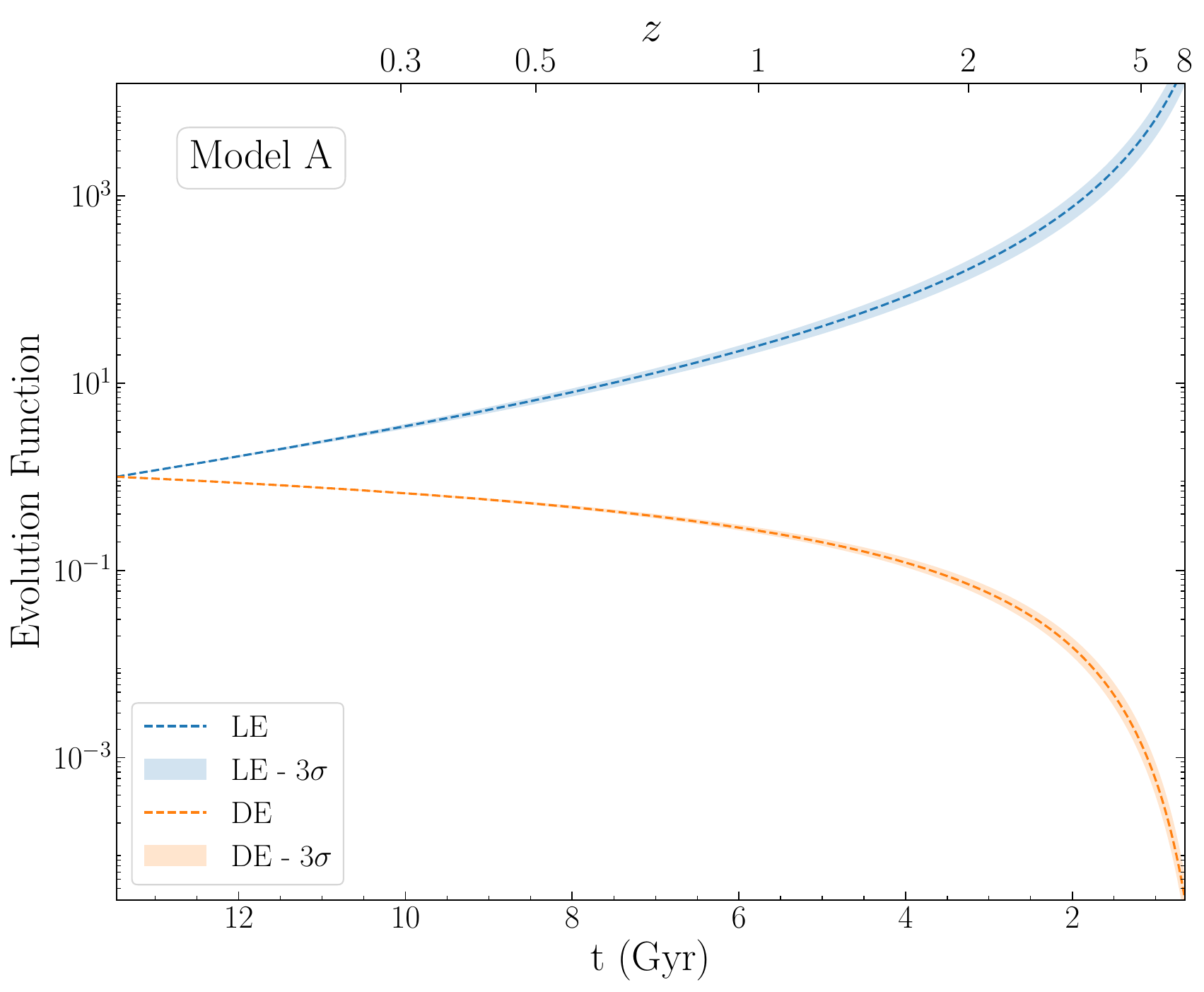}{0.33\textwidth}{}
        \hspace{-0.1in}
        \fig{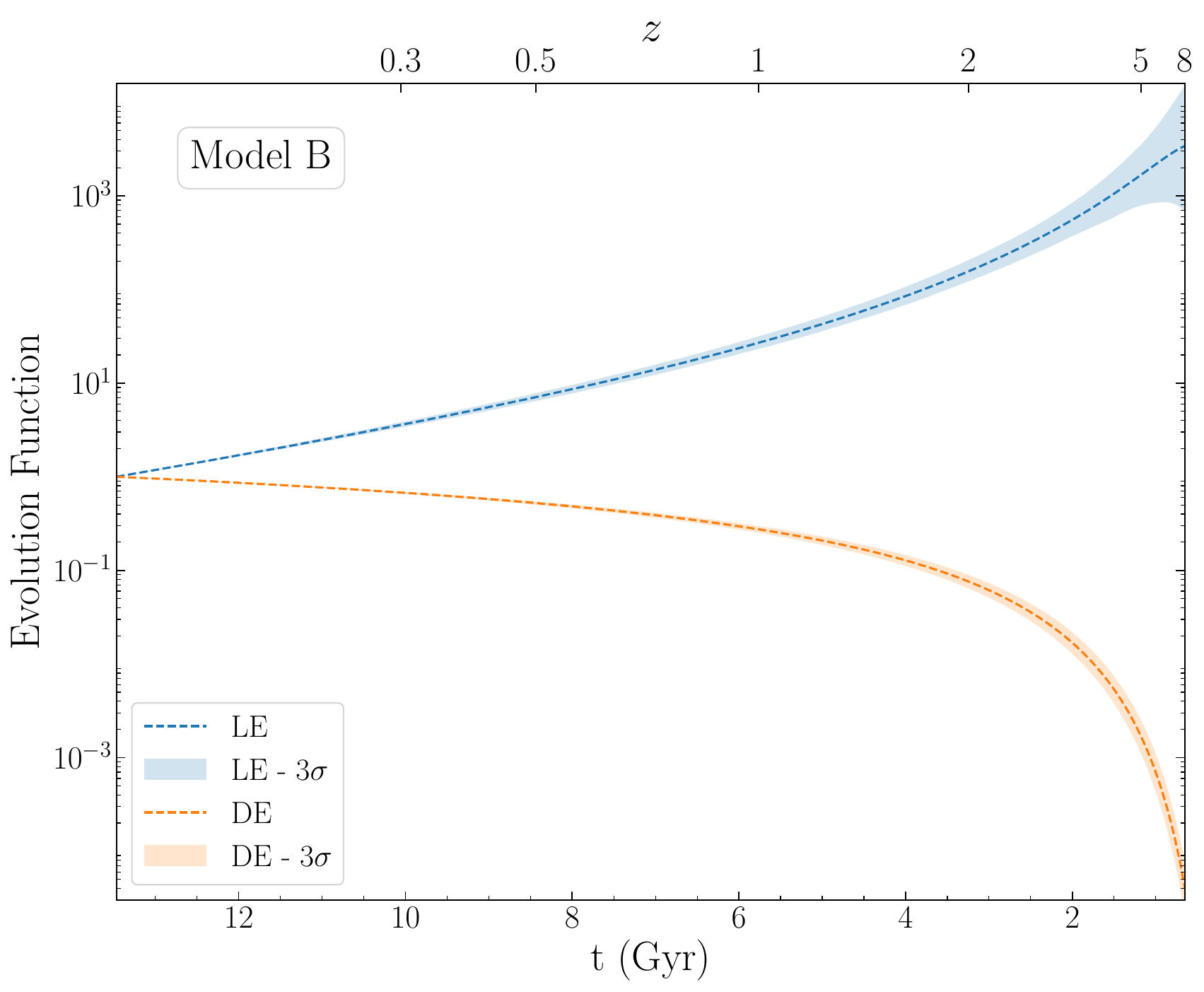}{0.33\textwidth}{}
        \hspace{-0.1in}
        \fig{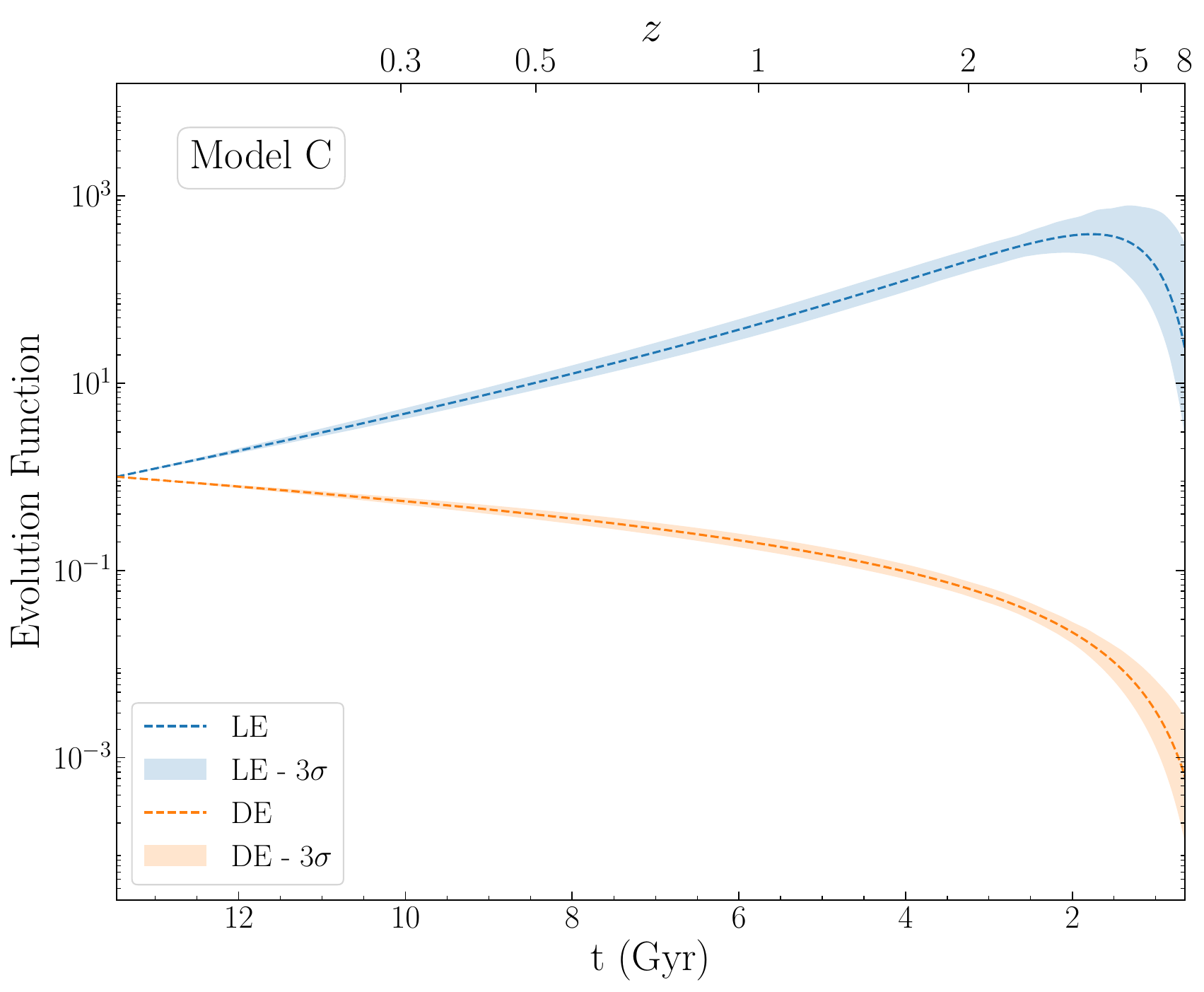}{0.33\textwidth}{}
    }
    \vspace{-0.3in} 
    \caption{
    Evolution of the density (orange) and luminosity (blue) functions for Models~A, B, and C. Top panels show the evolution as a function of redshift ($z$), while
    bottom panels display the evolution as a function of cosmic age ($t$, in Gyr).
    The lightly shaded regions indicate the $3\sigma$ confidence intervals.
    }
    \label{fig:DELE}
\end{figure*}

\subsection{Comparison with Observed Source Counts}

To place our model predictions in a broader observational context, we compare the source counts predicted by our parametric RLF models with both AGN-classified source-count datasets and total radio source-count measurements.
We calculate the SCs for each of our three LADE model LFs using Equation (\ref{eq:sc1}).
The results are presented in Figure \ref{fig:Source_Counts}, where the blue, orange and green dash-dotted lines correspond to Models A, B, and C, respectively.
The AGN-classified Euclidean-normalized $1.4$\,GHz source counts shown in this figure are compiled from \citet{Padovani2015}, \citet{Smolcic2017a}, \citet{Algera_2020}, and the AGN source-count measurements of \citet{2023MNRAS.520.2668H}. We also show total radio source counts from \citet{2010A&ARv..18....1D}, \citet{Algera_2020}, and the total source-count measurements of \citet{2023MNRAS.520.2668H} for comparison.

The AGN-classified source counts are the quantities relevant for comparison with our steep-spectrum AGN RLF model, whereas the total radio source counts are shown only to illustrate how the overall radio population increasingly exceeds the AGN-only counts toward the faint end. In particular, the AGN-classified source-count datasets (together with the bright-end AGN-dominated counts from \citet{2010A&ARv..18....1D} above $1\,\mathrm{mJy}$) are also included as external constraints in the joint fit through Equation~(\ref{chi2all}). Therefore, the comparison with the AGN source counts should be interpreted primarily as showing how well the fitted models reproduce the imposed AGN source-count constraints, rather than as a fully independent validation.

All three models reproduce the AGN-classified source counts reasonably well in the bright regime ($F_\nu > 1\,\mathrm{mJy}$). Toward lower flux densities, the total radio source counts lie systematically above the AGN-classified counts, illustrating the growing importance of non-AGN radio populations in the full counts. The AGN-classified counts themselves also tend to remain above our model predictions at the faint end. This residual difference should be interpreted cautiously: it may reflect contributions from AGN populations not explicitly included in our model, such as radio-quiet AGN, as well as differences in source-classification schemes across the literature. Since our model is intended to describe only the steep-spectrum radio-loud AGN population, we do not expect an exact match to either the total radio source counts or to all faint-end AGN-classified counts.

\subsection{Density and Luminosity Evolution}
\label{sec:DE_LE}

Figure 6 summarizes the inferred evolution of the DE and LE functions for Models A--C. The upper panels display these evolutionary trends as a function of redshift ($z$), while the lower panels present them as a function of cosmic age ($t$). In the context of our LADE framework, these two components play distinct physical roles: the LE function, $e_2(z)$, describes the evolution of the characteristic luminosity scale (shifting the ``knee'' of the luminosity function, $L_* \to L_* e_2(z)$), whereas the DE function, $e_1(z)$, governs the evolution of the global normalization ($\phi_*$) at a fixed luminosity.

As observed in the upper panels, across all three models, the LE term exhibits a strong positive evolution from $z \simeq 0$ to the ``cosmic noon'' epoch ($z \sim 2$), indicating a rapid increase in the characteristic radio power of steep-spectrum AGNs toward earlier cosmic times. Model A continues a monotonic rise by construction because it adopts a simple power-law form for $e_2(z)$. However, Models B and C, which allow for redshift-dependent indices, reveal a saturation or mild turnover in the characteristic luminosity at $z \ge 3.5$. This suggests that the typical luminosity of radio AGNs does not grow indefinitely but likely plateaus during the epoch of initial supermassive black hole assembly.

In contrast to the rising LE, all models favor a negative DE with increasing redshift, indicating a systematic decline in the LF normalization toward earlier epochs. For Models A and B, this decrease is explicitly exponential in $z$ via the adopted $e_1(z) = \exp(-p_1 z)$ form, while Model C permits additional flexibility and shows a slower decline  at $z \ge 2.5$. Consequently, the evolution of steep-spectrum AGNs is best described by a ``dual'' process: a decrease in the overall number density of sources, counterbalanced by a significant increase in their intrinsic luminosities.

We emphasize that while DE and LE are mathematically distinct in Equation (10), their physical decoupling provides key insights into AGN growth. The monotonic decline of $e_1(z)$ implies that the total comoving number density of steep-spectrum AGNs decreases toward higher redshift. Meanwhile, the rapid rise of $e_2(z)$ up to $z \sim 2-3$ indicates that although these sources are rarer in the early Universe, they are systematically more powerful. This is consistent with a scenario where high-redshift radio galaxies represent a phase of highly efficient fueling and rapid black hole growth, coinciding with the peak of cosmic star formation activity.

We stress that the DE/LE trends inferred here describe an effective global evolution of the full steep-spectrum radio-loud AGN sample treated as a single statistical population. This does not imply that  radio-loud AGNs arise from a single physical accretion/feedback mode, nor does it rule out an underlying bimodality (e.g. high-/low-excitation radio galaxies or cold-/hot-mode accretion). Rather, our results indicate that, for the radio-only LF of the steep-spectrum (FSRQ-excluded) sample, the data do not require two independent evolutionary tracks: any distinct physical modes may substantially overlap in radio luminosity and redshift, or their relative contributions may vary smoothly with $(z,L)$, yielding the observed smooth DE and LE behavior. A direct test of physical bimodality would require modelling the LFs separately for physically classified subsamples, which we defer to future work.

\begin{figure}
	\centering
	\includegraphics[width=0.33\textwidth]{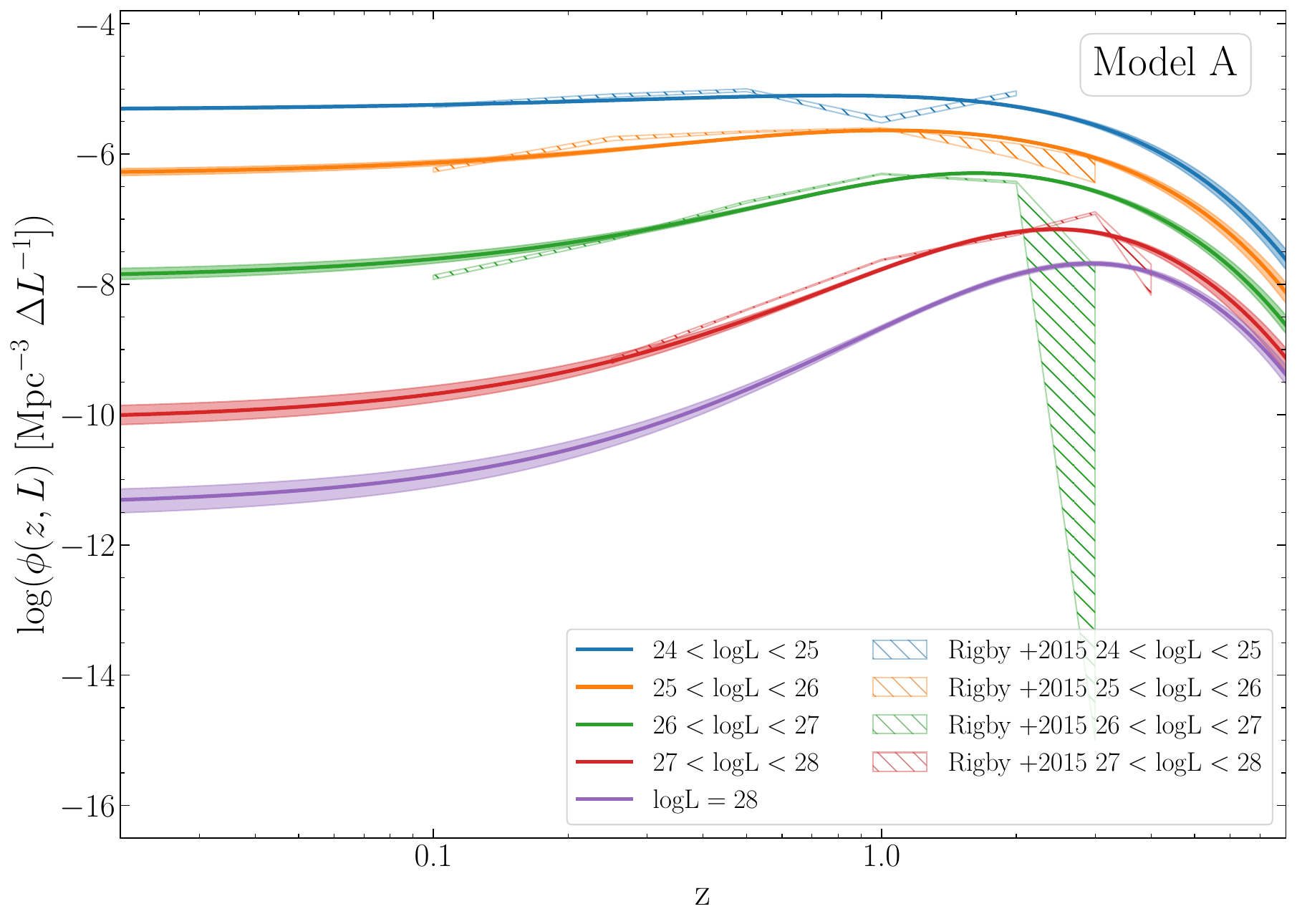}
	\hspace{-0.1in}
	\includegraphics[width=0.33\textwidth]{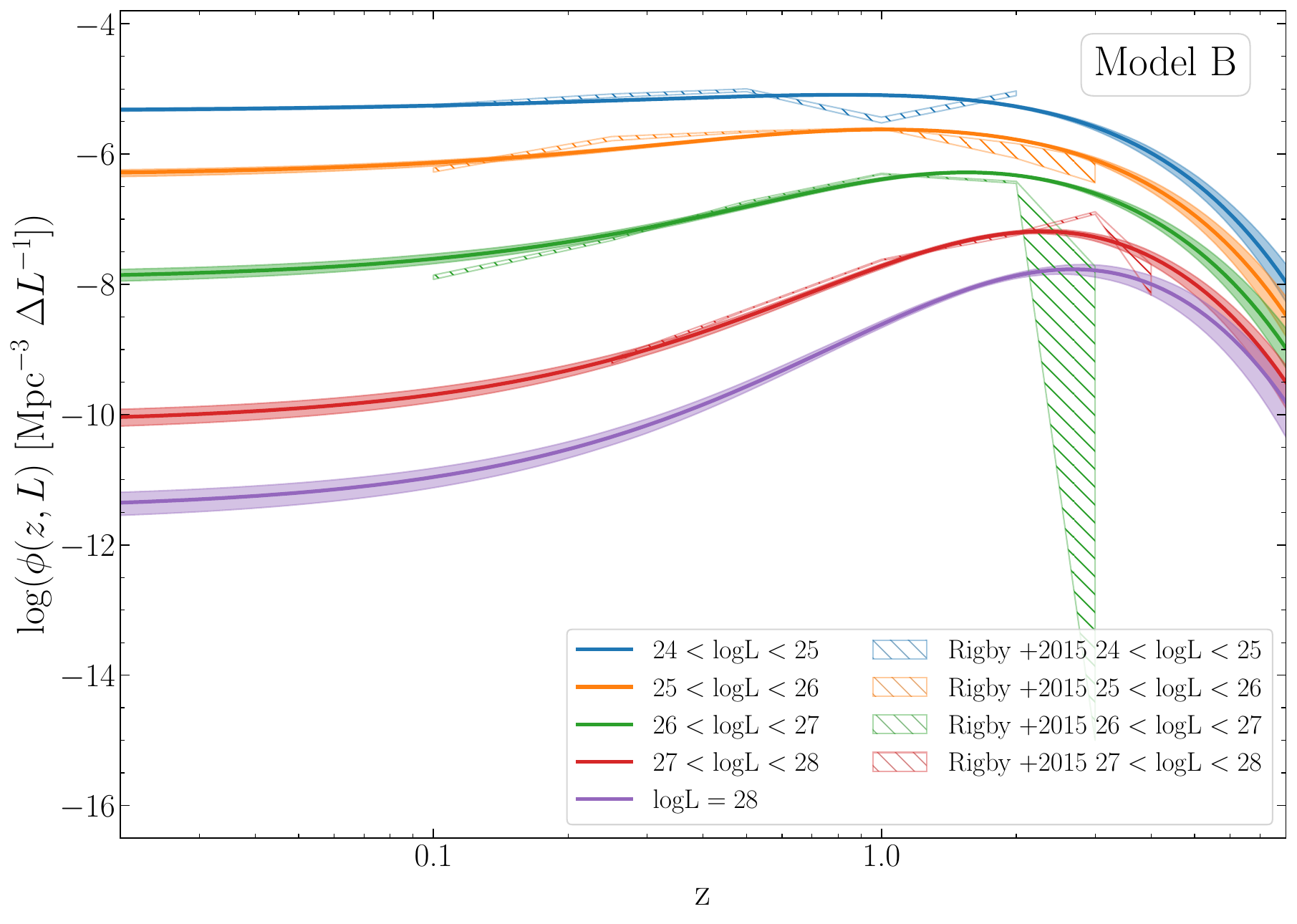}
    \hspace{-0.1in}
	\includegraphics[width=0.33\textwidth]{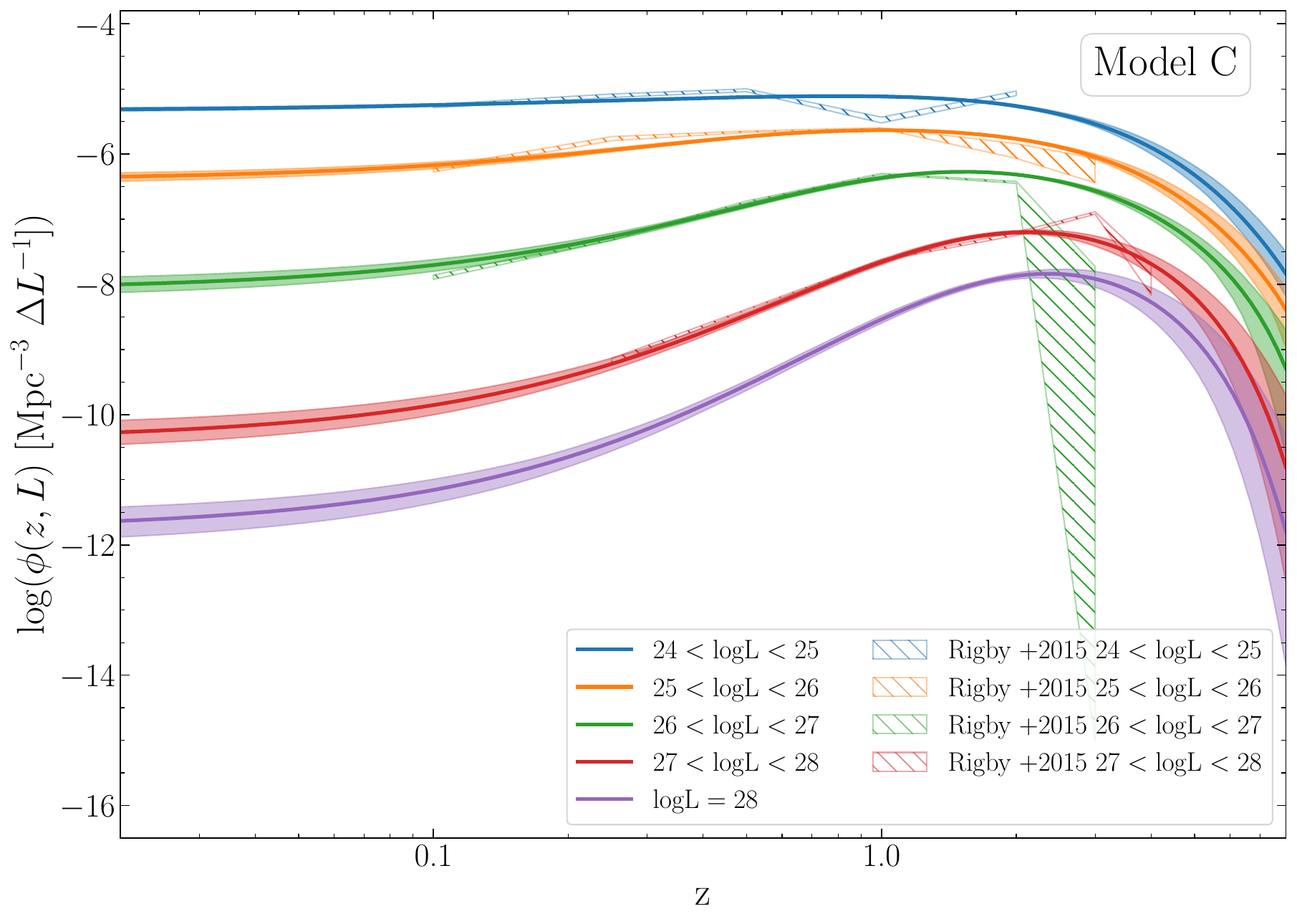}
	\caption{Redshift evolution of the comoving space density for Models A, B, and C.
    The blue, orange, green, red, and purple solid lines represent the radio RLFs at $\log(L_{1.4\,\mathrm{GHz}} [\mathrm{W\,Hz^{-1}}]) = 24.5$, 25.5, 26.5, 27.5, and 28.0, respectively.
    The shaded regions indicate the $3\sigma$ uncertainty ranges.
    For comparison, the results from \citet{Rigby2015} are shown as blue, orange, green, and red hatched regions.}
	\label{fig:LF_Lum}
\end{figure}

\begin{figure*}
    \centering
    \gridline{
        \fig{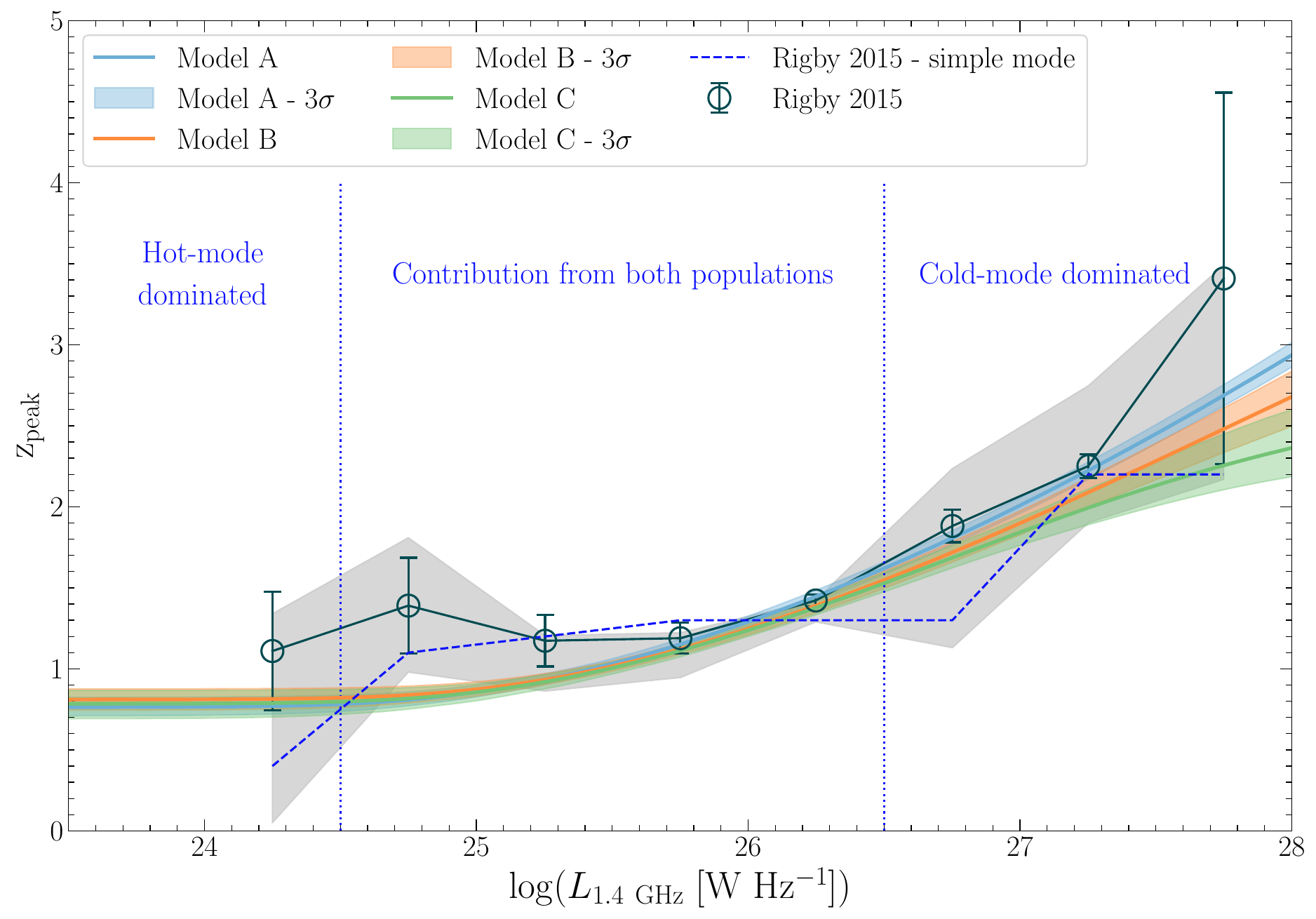}{0.5\textwidth}{(a)}
        \hspace{-0.1in}
        \fig{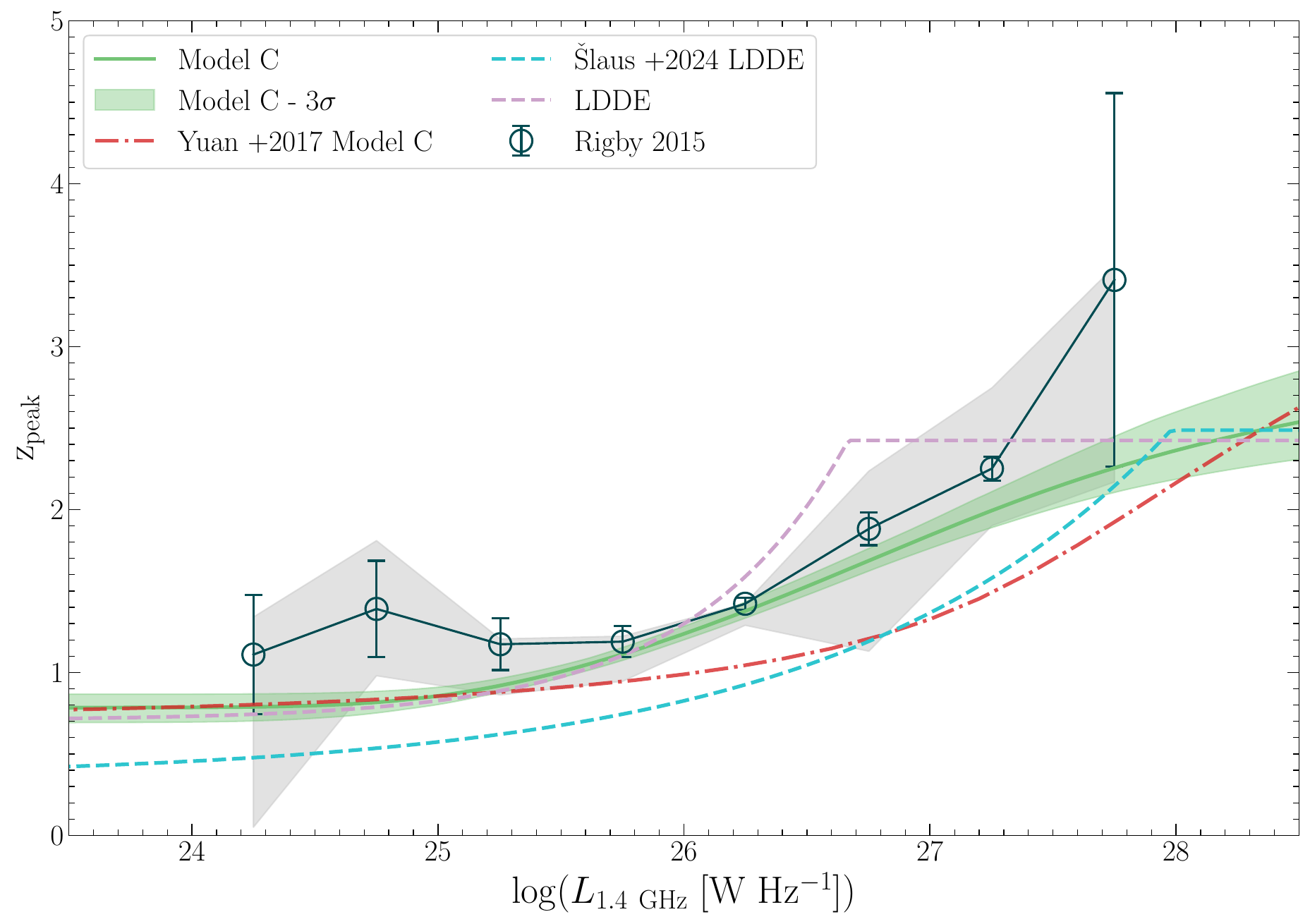}{0.5\textwidth}{(b)}
    }
    \vspace{0.01in}
    \caption{Variation in the redshift of the peak space density with radio luminosity and comparisons with literature results. 
    \textbf{(a)} The blue, orange, and green dashed lines represent our Models A, B, and C, respectively, with shaded regions indicating the $3\sigma$ confidence intervals. The red dash-dotted line shows the prediction from Model C of \citet{yuan2017mixture}, while the dark blue circles with error bars denote the results from \citet{Rigby2015}. The grey shaded area illustrates the range in their results found by varying the input parameters in the RLF grid modelling. The blue dashed line corresponds to the simple evolutionary model proposed in \citet{Rigby2015}.
    \textbf{(b)} The green solid line indicates our Model C, with the surrounding shaded area marking its $3\sigma$ uncertainty. The red dotted line represents Model C from \citet{yuan2017mixture}, and the circular markers with error bars show the R15 measurements. The light blue dashed line corresponds to the LDDE model from \citet{2024A&A...684A..19S}, while the purple dashed line shows the LDDE model in this work, both with shaded regions indicating their respective $3\sigma$ confidence intervals.
    }
    \label{fig:zpeak}
\end{figure*}

\subsection{Luminosity-dependent Evolution}

As discussed by \citet{Rigby2015} and \citet{2016ApJ...829...95Y,yuan2017mixture}, the position of the steep-spectrum RLF peak is luminosity dependent; specifically, the comoving space density of the most powerful sources peaks at higher redshift than that of their weaker counterparts.
Figure \ref{fig:LF_Lum} illustrates the redshift evolution of the comoving space density for Models A, B, and C at various luminosities. The inferred evolution exhibits a clear signature of ``cosmic downsizing'': the evolutionary trends are strongly dependent on radio luminosity. Moreover, the amplitude of the evolution from $z = 0$ to the redshift at which the space density peaks is also highly luminosity-dependent.
For high-luminosity sources, the variation in space density spans more than two orders of magnitude, whereas for low-luminosity sources, the change is less than an order of magnitude.
This behavior is consistent with what has been observed for quasars (QSOs) selected in the X-ray and optical bands \citep[e.g.,][]{2005A&A...441..417H, Hopkins2007}.

For comparison, we also show the results from \citet{Rigby2015} in Figure~\ref{fig:LF_Lum}.
Our model predictions are in good agreement with \citet{Rigby2015}.
However, it is worth noting that in the luminosity range $24 < \log L < 25$, the radio LF from \citet{Rigby2015} exhibits a noticeable dip around $z \sim 1$.
This feature is unlikely to represent a real decline in space density; rather, it may be an artifact arising from the sensitivity of binned estimators to bin placement and widths.
As pointed out by \citet{Yuan2013}, such binning-based estimates are highly sensitive to the choice of bin edges and widths.

\subsection{Luminosity-dependent $z_{\rm peak}$}

Figure~\ref{fig:zpeak} shows the redshift at which the total comoving space density at fixed luminosity reaches its maximum, $z_{\rm peak}$, as a function of radio luminosity. In panel (a), we compare our three LADE models with the reference result of \citet{Rigby2015}. The blue, green, and orange dashed lines represent the predictions from Models~A, B, and C, respectively, with shaded regions indicating the corresponding $3\sigma$ confidence intervals. The dark blue open circles with error bars show the nonparametric results from \citet{Rigby2015}. Given the grid-based modeling approach adopted in that work (see original paper for details), the uncertainties arising from variations in input parameters are indicated by the grey shaded area. Our model predictions lie almost entirely within this region, demonstrating excellent consistency between our measurements and those of \citet{Rigby2015}. Furthermore, the overall trend in $z_{\rm peak}$ predicted by our models agrees very well with their results.

We also plot the simple evolutionary model proposed by \citet{Rigby2015} as the blue dashed line.
This model assumes two dominant accretion modes in radio galaxies, often discussed in terms of cold-mode/high-excitation and hot-mode/low-excitation accretion \citep[e.g.,][]{2007MNRAS.376.1849H,2012MNRAS.421.1569B,Heckman2014}. Cold-mode sources, typically high-luminosity AGNs, are associated with radiatively efficient accretion of cold gas (e.g., potentially triggered by galaxy mergers or interactions) and exhibit strong cosmic evolution. Their comoving density rises rapidly, peaking at a luminosity-dependent high redshift ($z \sim 2\text{--}4$), before declining sharply \citep[e.g.,][]{2007MNRAS.376.1849H}. Hot-mode sources, by contrast, are associated with radiatively inefficient accretion from hot halo gas. They show a much milder evolutionary trend with a weaker luminosity dependence, peaking in space density at a lower redshift around $z \sim 0.8$ \citep[e.g.,][]{2012MNRAS.421.1569B,Heckman2014}. Following the simple evolutionary picture of \citet{Rigby2015}, the total evolution is constructed by summing these two components at each luminosity. Therefore, we define $z_{\rm peak}(L)$ using this summed density rather than the individual components, which peak at different redshifts.
Remarkably, our LADE model predictions closely track the trajectory of this dual-mode scenario. This suggests that the LADE framework can phenomenologically capture the transition from hot-mode to cold-mode dominance with a single parametric form, without requiring an explicit decomposition into two populations.

In panel~(b), we adopt our preferred LADE Model~C as the baseline and compare it with other published determinations. The red dash-dotted line shows Model~C from \citet{yuan2017mixture}, which also employs the LADE formalism. While the overall trend is consistent, an offset is apparent. We attribute this primarily to the substantial improvement in sample statistics and phase-space coverage achieved here: our sample is $\sim$4 times larger than that of \citet{yuan2017mixture}, and, crucially, the inclusion of the XXL and COSMOS surveys provides much tighter constraints on the faint end that were previously less well determined. As a result, the present $z_{\rm peak}(L)$ relation should be viewed as a robust refinement of the earlier estimates. In addition, to a lesser extent, systematic uncertainties arising from the conversion of the 408 MHz RLF to 1.4 GHz (assuming a fixed spectral index $\alpha = 0.75$) in \citet{yuan2017mixture} may also contribute to the observed offset.

Panel~(b) also compares our LADE result with the LDDE-based predictions. The light blue dashed line shows the LDDE result from \citet{2024A&A...684A..19S}, while the purple dashed line corresponds to the LDDE fit obtained in this work.
Crucially, neither of these LDDE parameterizations reproduces the non-parametric results of \citet{Rigby2015} as faithfully as our LADE model. Moreover, the LDDE forms exhibit an artificial feature where the rise of $z_{\rm peak}$ abruptly saturates into a flat plateau at high luminosities. In contrast, the LADE model yields a smooth, continuous dependence that appears more physically plausible.

\begin{figure*}[ht]
    \centering
    \gridline{
        \fig{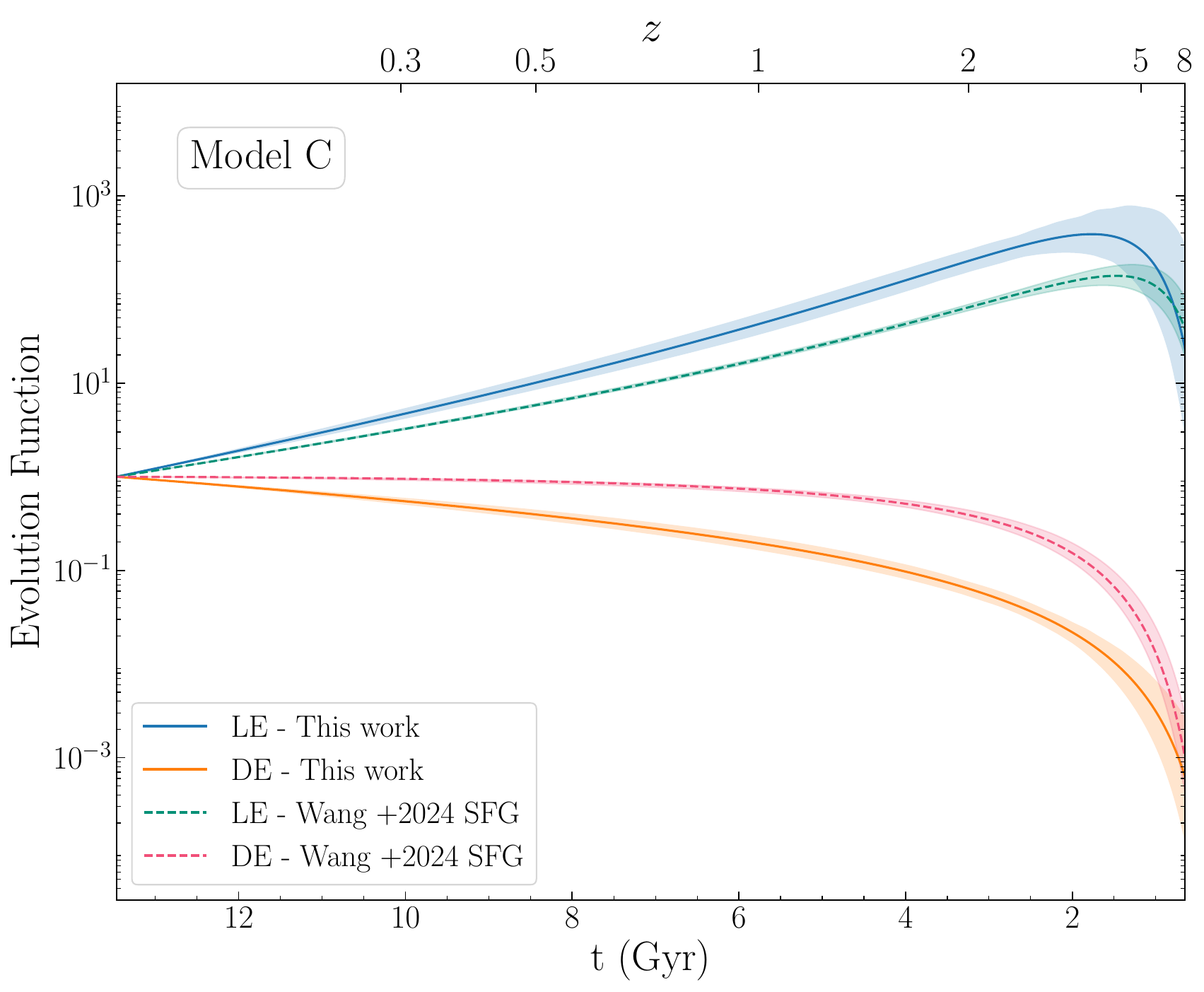}{0.33\textwidth}{(a) Evolution function}
        \hspace{-0.1in}
        \fig{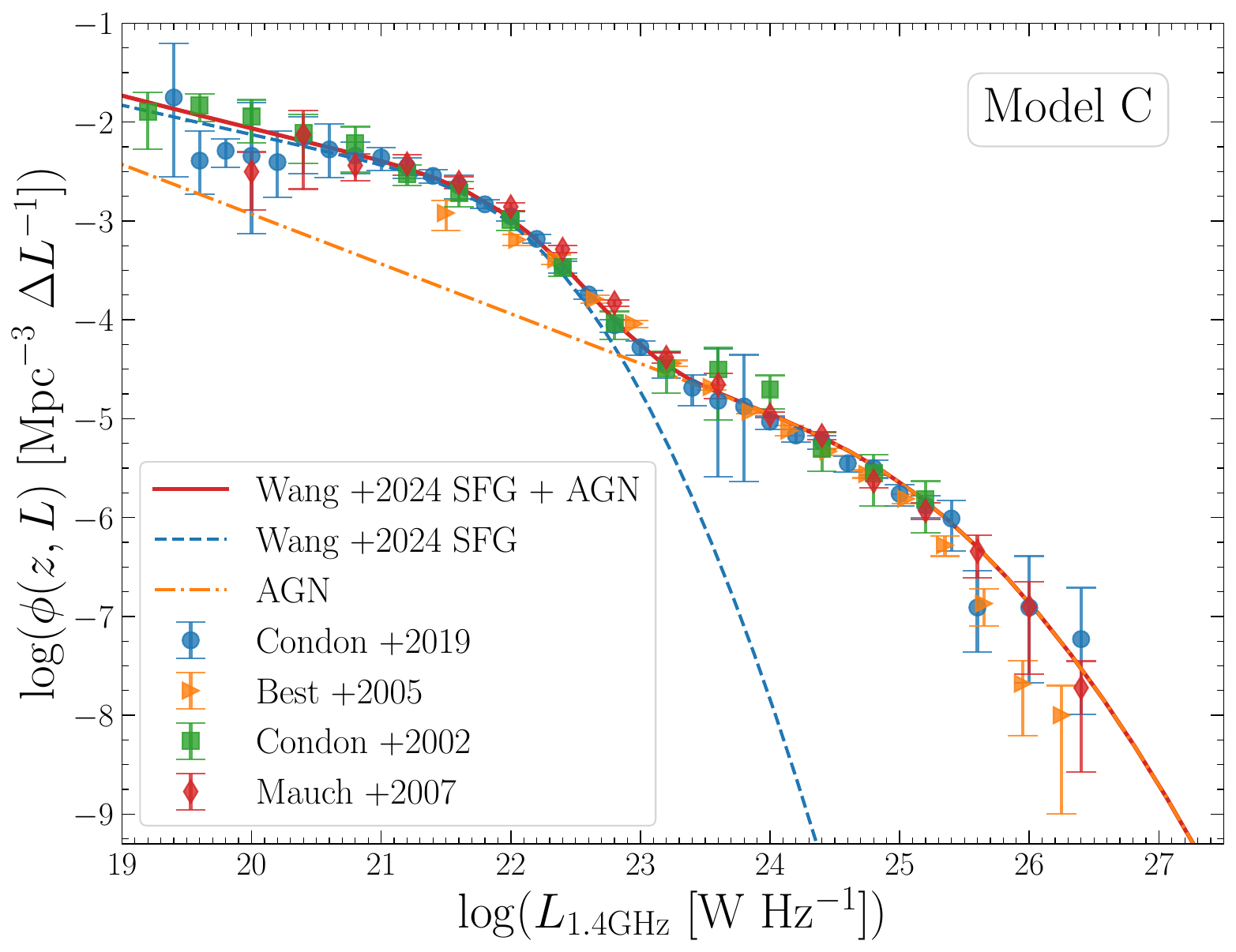}{0.33\textwidth}{(b) Local LF}
        \hspace{-0.1in}
        \fig{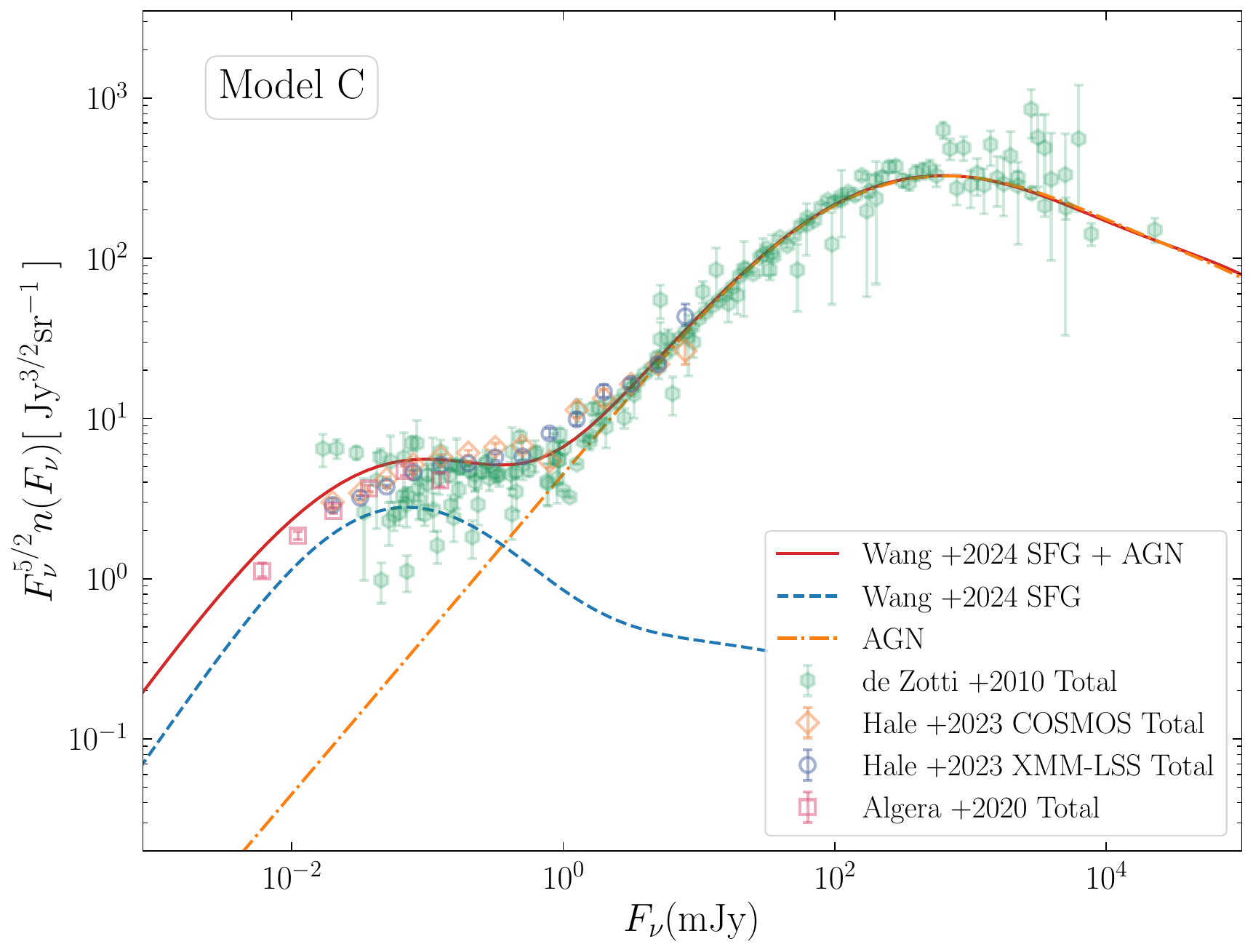}{0.33\textwidth}{(c) Source count}}  
    \vspace{0.01in}
\caption{Unified two-component description of the 1.4~GHz radio population under Model~C. (a) Evolution functions versus cosmic age $t$. Solid lines: AGN DE (blue) and LE (orange). Dashed lines: SFG DE (green) and LE (pink) from W24. Shading shows the $3\sigma$ confidence intervals. (b) Local ($z=0.1$) RLFs for SFGs (green dashed) and AGNs (orange dash-dotted) and their sum (total; red solid), compared to the local measurements of \citet{Condon_2019}, \citet{best2005sample}, \citet{Condon_2002}, and \citet{Mauch_2007}. (c) Euclidean-normalized differential source counts at 1.4~GHz for SFGs (green dashed) and AGNs (orange dash-dotted) and their sum (total; red solid), compared with the bright-end radio source count compilation of \citet{2010A&ARv..18....1D}, the COSMOS/XMM-LSS counts of \citet{2023MNRAS.520.2668H}, and the measurements of \citet{Algera_2020}.}
    \label{fig:SFG_AGN_C}
\end{figure*}

\begin{figure*}
    \centering
    \gridline{
        \fig{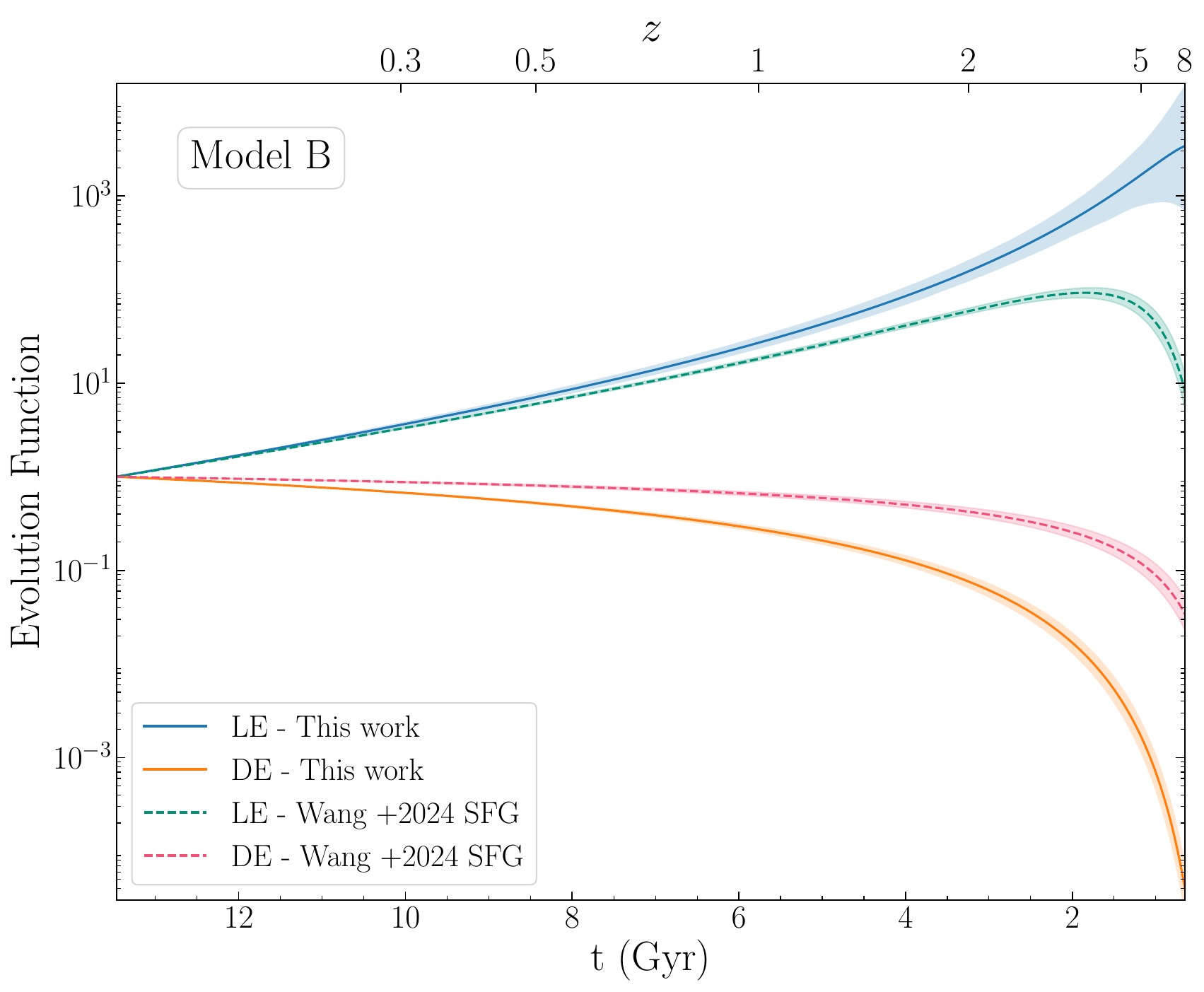}{0.33\textwidth}{(a) Evolution function}
        \hspace{-0.1in}
        \fig{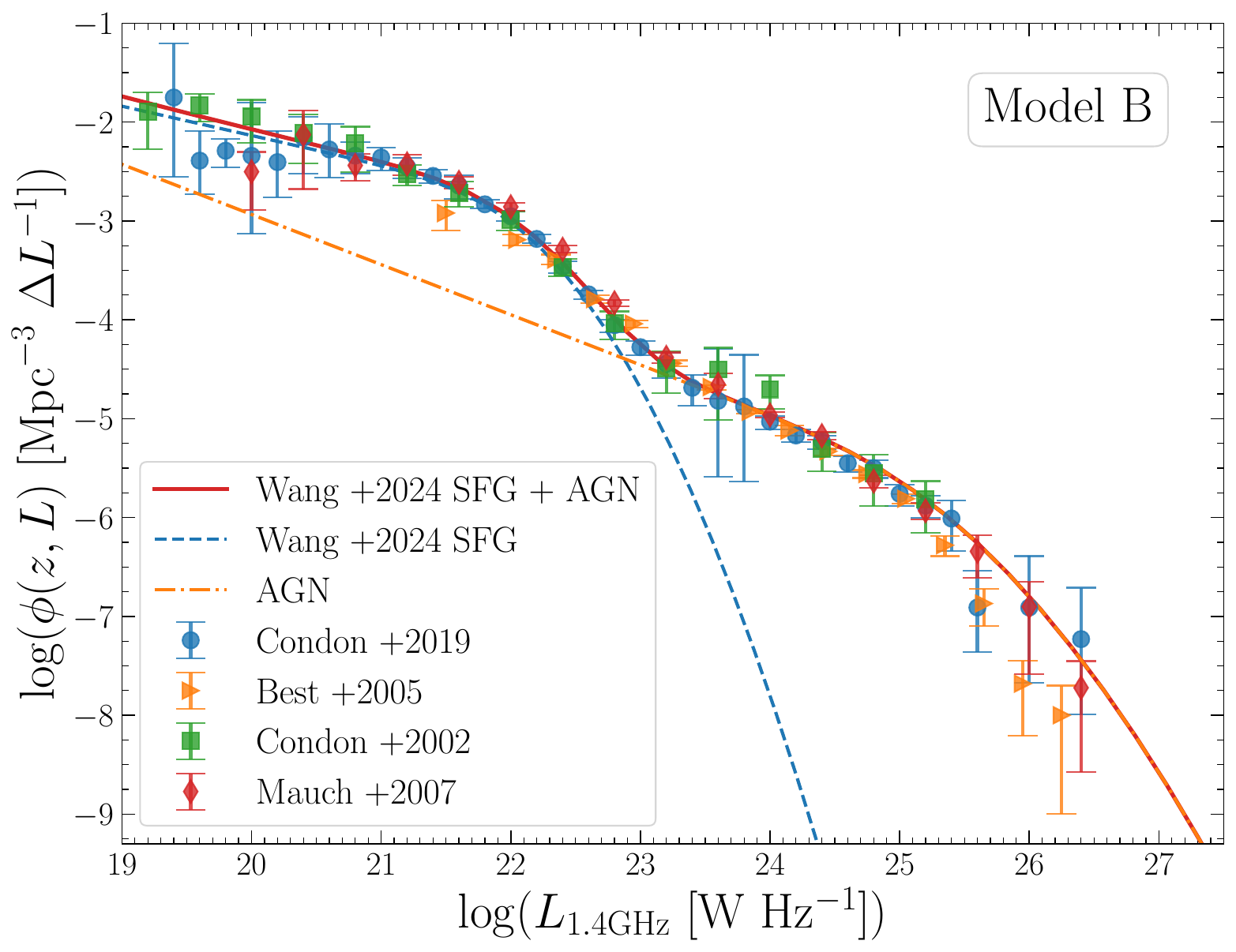}{0.33\textwidth}{(b) Local LF}
        \hspace{-0.1in}
        \fig{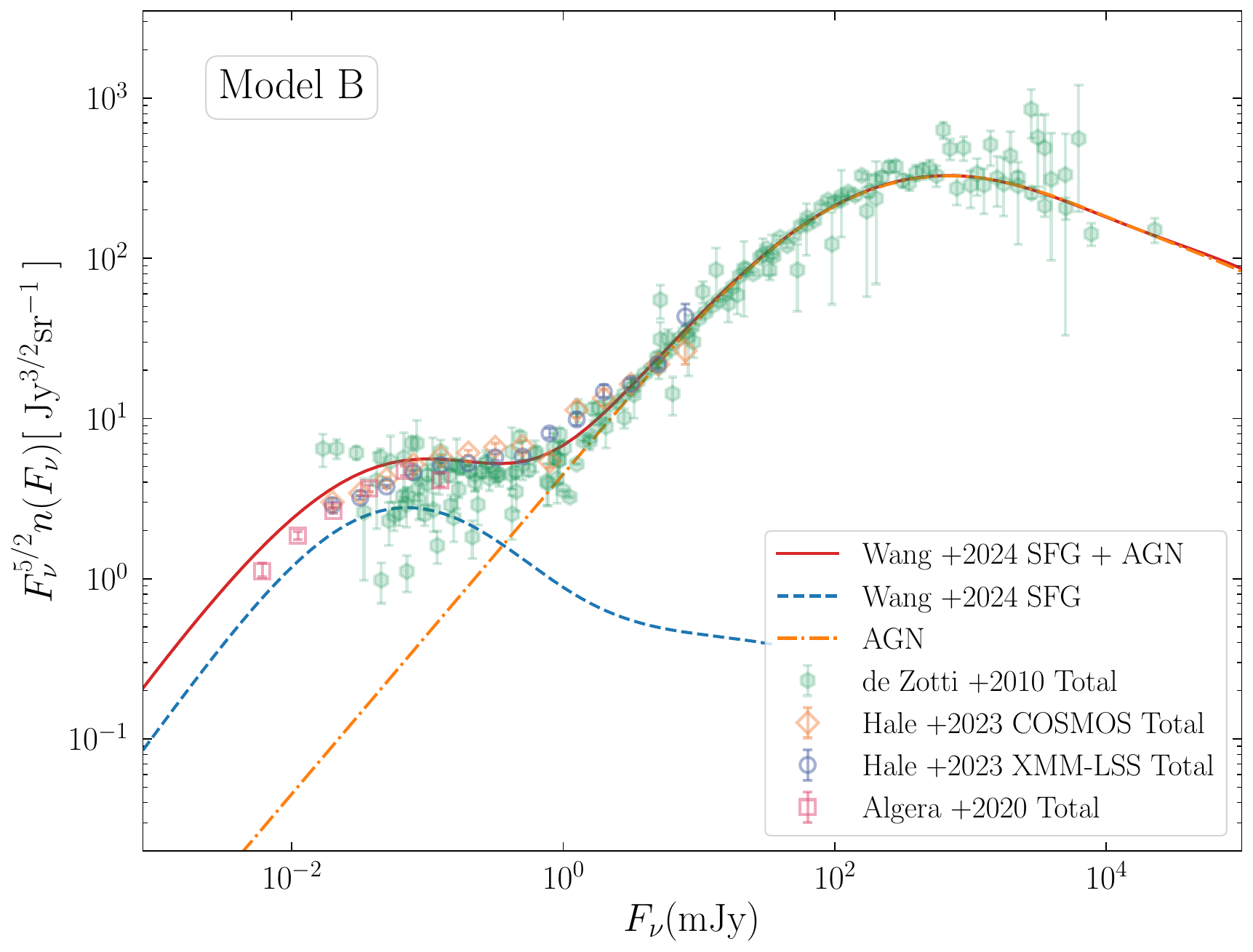}{0.33\textwidth}{(c) Source count}}
    \vspace{0.01in}
    \caption{Same as Figure~\ref{fig:SFG_AGN_C}, but for Model~B.}
    \label{fig:SFG_AGN_B}
\end{figure*}

\section{DISCUSSION}
\label{sec_discussion}

\subsection{A Unified Evolutionary Framework for AGNs and SFGs}
\label{sec:SFG_AGN_coevo}
Having established a LADE-based description of the radio-loud AGN LF, we now cast it in the same parametric framework used to model the SFG radio LF in our previous work \citep[][hereafter W24]{2024A&A...683A.174W}. Although the AGN and SFG analyses were conducted independently, the preferred prescriptions identified in the two studies—Models B and C for AGNs and the W24 model for SFGs—converge on an identical mathematical architecture. In particular, they adopt the same family of local LFs and employ the same functional forms for DE and LE.

Importantly, this concordance was not enforced a priori. In fitting the AGN sample, we tested a broad suite of candidate DE/LE parameterizations and retained Models B and C solely on statistical grounds, i.e., their favorable balance between goodness-of-fit and parsimony as quantified by information criteria (e.g., AIC). That the resulting AGN models coincide with the SFG prescriptions from W24 therefore emerges a posteriori, providing a rigorous and standardized ``common language'' for controlled comparisons between these physically distinct populations.

We next focus on our fiducial unified description under Model~C, presented in Figure~\ref{fig:SFG_AGN_C}. Panel~(a) compares the best-fitting DE and LE terms for the SFG component from W24 and the steep-spectrum AGN component derived here. Despite the fact that the two populations are governed by distinct parameter values, their inferred LE histories are broadly similar: the characteristic luminosities rise toward a maximum at $z\sim2$--3 and then flatten or exhibit a mild decline at higher redshift. By contrast, the DE terms decrease with redshift for both populations, remaining monotonic over the redshift range probed. Overall, the two components display qualitatively consistent evolutionary trends within the same LADE functional family, while retaining population-specific amplitudes and detailed redshift dependences.

A closer comparison, however, reveals systematic differences in the strength of the evolution. In the rising phase toward cosmic noon, the AGN LE increases more steeply than that of SFGs, implying a more rapid brightening of the characteristic radio power with redshift. This behavior is qualitatively consistent with the expectation that the processes governing radio-loud AGN activity---e.g., the availability of fuel for SMBH accretion and the triggering of powerful jets---are more strongly peaked in time around $z\sim2$--3 than the mechanisms regulating star formation in the general galaxy population, which evolve more smoothly.

The contrast is even more pronounced in the DE term: the AGN DE declines faster with redshift than the SFG DE over the same interval. Within the LADE decomposition, this indicates that the comoving abundance (or, equivalently, the effective duty cycle) of the radio-loud AGN population decreases more rapidly toward earlier cosmic times than that of SFGs. Taken together, the steeper LE rise and the faster DE decline suggest that, although both populations can be described within a unified functional framework, radio-loud AGNs exhibit a stronger redshift dependence---i.e., a more pronounced cosmological evolution---than SFGs in Model~C.

Panel~(b) of Figure~\ref{fig:SFG_AGN_C} shows the implied local ($z=0.1$) 1.4~GHz RLFs for the two components and their sum. Specifically, we combine the steep-spectrum AGN LF inferred under our Model~C with the SFG LF from W24 (also under Model~C) to obtain the total RLF (red solid curve). The resulting total RLF agrees well with a range of independent local measurements, including those from \citet{Condon_2019}, \citet{best2005sample}, \citet{Condon_2002}, and \citet{Mauch_2007}, indicating that the unified decomposition provides a consistent description of both the faint and bright ends of the local radio population. Panel~(c) presents an analogous comparison in terms of the Euclidean-normalized differential source counts. Here we sum the SFG and AGN contributions predicted by the unified Model~C to obtain the total 1.4~GHz radio counts, and compare them with observational determinations from \citet{2010A&ARv..18....1D}, \citet{2023MNRAS.520.2668H} and \citet{Algera_2020}. 
We emphasize that Panel~(c) of Figure~\ref{fig:SFG_AGN_C} serves a different purpose from Figure~\ref{fig:Source_Counts}: it illustrates the global two-component (SFG+AGN) decomposition of the total radio source counts, whereas the source counts used in Equation~(\ref{chi2all}) are AGN-classified counts only.
The combined model reproduces both the overall normalization and the flux-dependent curvature of the observed counts over the full range probed, indicating that the same two-component decomposition that matches the local LF also provides a self-consistent description of the global source-count statistics.

Figure~\ref{fig:SFG_AGN_B} presents the same unified, two-component decomposition as Figure~\ref{fig:SFG_AGN_C}, but adopting Model~B. The overall behavior is unchanged: the AGN and SFG evolution functions remain broadly consistent within the shared LADE architecture, their summed local ($z=0.1$) RLF continues to match the published $z\approx0$ measurements, and the combined Euclidean-normalized source counts reproduce the observed normalization and curvature over the relevant flux-density range. Consequently, the main qualitative trends inferred from Model C remain unchanged when adopting Model B, indicating that our conclusions are not sensitive to the choice between these two preferred parameterizations. We nevertheless adopt Model C as our fiducial model because it yields slightly better information-criterion values.

The similarity of the inferred LE and DE trends for SFGs and radio-loud AGNs provides a useful population-level baseline for co-evolution studies, while not by itself implying a direct, galaxy-by-galaxy causal link between star formation and accretion.

\begin{figure*}
    \centering
    \includegraphics[width=0.6\textwidth]{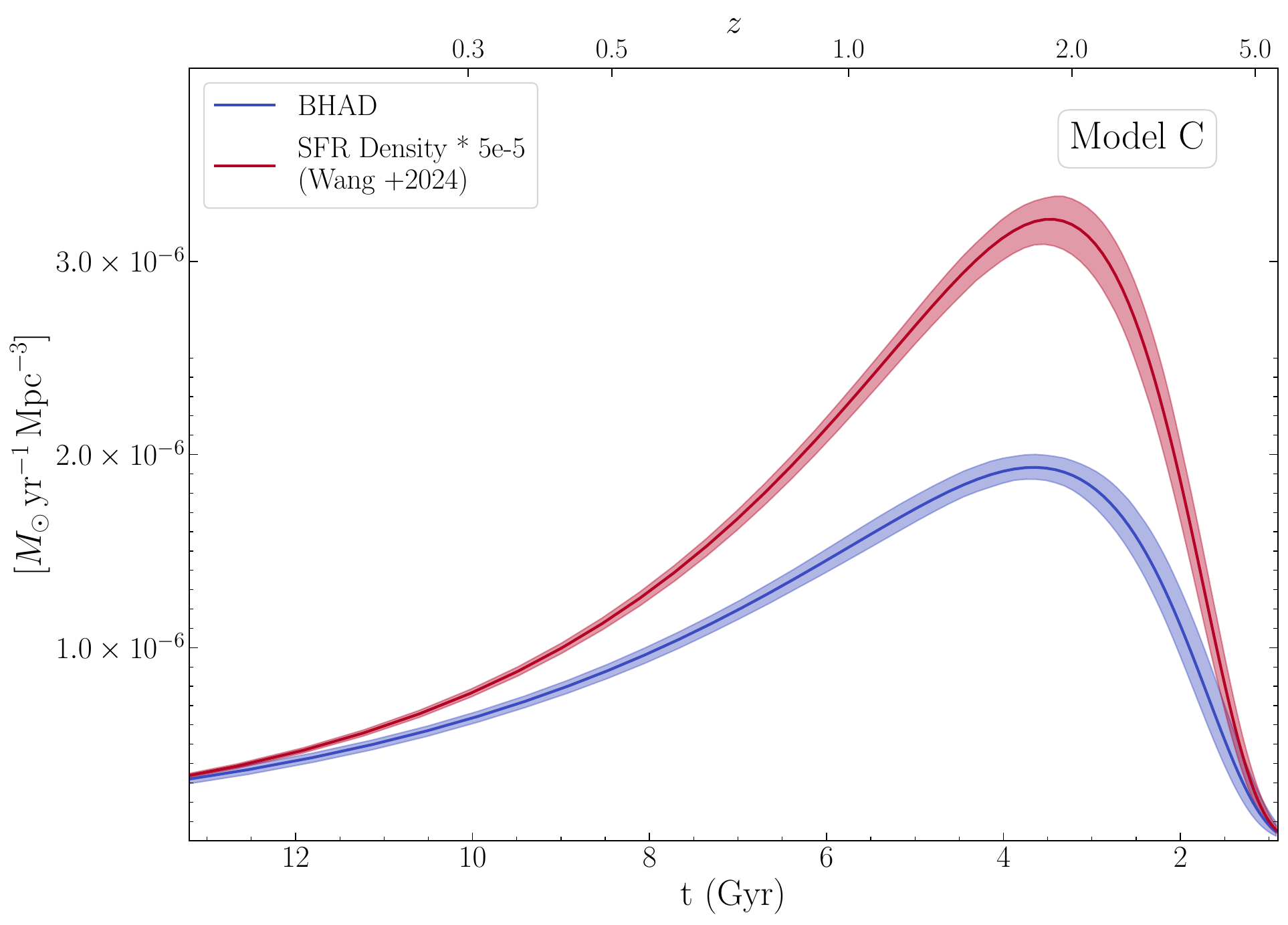}
    \caption{
    Cosmic evolution of the BHAD (blue) compared with the SFRD (red) derived from our previous work \citep{2024A&A...683A.174W}. The SFRD has been rescaled by a factor of $5\times10^{-5}$ to enable direct comparison with the BHAD. Solid curves denote the best-fit predictions, while shaded regions indicate the associated $3\sigma$ uncertainties.  
    }
    \label{fig:bhar}
\end{figure*}

\subsection{Deriving the Black Hole Accretion Rate Density}
\label{sec_bhar}

Having derived the rest-frame 1.4\,GHz LF of our steep-spectrum radio-loud AGN sample, we next estimate the corresponding radio-mode (kinetic) BHAD ($\Psi_{\rm bhar}$) over cosmic time, from $z\sim4$ down to the present epoch. We stress that this quantity does not represent the total cosmic BHAD, since our analysis excludes radio-quiet AGN and also omits the flat-spectrum radio-loud population removed in Section~\ref{sub_fsrq} to minimize beaming-related biases.
Following \citet{2024A&A...684A..19S} \citep[see also][]{Smolcic2017c,2018A&A...620A.192C}, we determined the kinetic luminosity from the radio luminosity using the relation
\begin{equation}
\label{eq:Lkin_Lrad}  
\log L_{\mathrm{kin}}=0.86\log L_{1.4~\mathrm{GHz}}+14.08+1.5\log f_{\mathrm{W}},
\end{equation}
where $L_{\mathrm{kin}}$ is the kinetic luminosity, $L_{1.4 \,\mathrm{GHz}}$ is the rest-frame 1.4\,GHz radio luminosity, and $f_{\mathrm{W}}$ is a parameter introduced by \citet{Willott1999} to account for all possible systematic uncertainties.
Based on observational constraints, $f_{\mathrm{W}}$ is typically estimated to lie in the range $1$--$20$.
We assumed $f_{\mathrm{W}}=15$, for which the \citet{Willott1999} relation is consistent with jet kinetic luminosities inferred from X-ray cavity observations in galaxy groups and clusters \citep{2004ApJ...607..800B,2011ApJ...735...11O,Smolcic2017c,2018A&A...620A.192C,2024A&A...684A..19S};
for a fixed adopted value of $f_{\mathrm{W}}$, this acts primarily as a multiplicative normalization shift in the kinetic luminosity density, without by itself changing the redshift dependence. 
We emphasize, however, that this radio-to-kinetic conversion remains one of the dominant systematic uncertainties in our BHAD calculation. The \citet{Willott1999} relation is an analytical scaling derived from minimum-energy arguments for radio lobes, and the factor $f_{\mathrm W}$ is intended to absorb uncertainties related to the lobe composition, filling factor, departure from equipartition, low-frequency cutoff, and other aspects of jet physics. In addition, the empirical calibration between monochromatic radio luminosity and kinetic jet power shows substantial intrinsic scatter, and may itself depend on environment and redshift. Similar caveats apply to cavity-power-based calibrations. We therefore interpret the kinetic luminosity density and the resulting BHAD primarily in a relative sense, with the redshift evolution likely being more robust than the absolute normalization.

Using Equation (\ref{eq:Lkin_Lrad}), we derive the kinetic LF as
\begin{equation}
\begin{aligned}
\label{eq:klf}
\Phi(L_{\rm kin},z)\mathrm{d}\log L_{\rm kin}
= \Phi_{1.4~\mathrm{GHz}}(L_{1.4~\mathrm{GHz}},z)\mathrm{d}\log L_{1.4~\mathrm{GHz}},
\end{aligned}
\end{equation}
where the transformation follows from the logarithmic relation in Equation (\ref{eq:Lkin_Lrad}).
The BHAD is then computed using a ``So\l{}tan-type'' argument:
\begin{equation}
\label{eq:bhad_kin}
\Psi_{\rm BHAR}(z)
=
\frac{1-\epsilon_{\rm rad}-\epsilon_{\rm kin}}
{\epsilon_{\rm kin}\,c^2}
\int \mathrm{d}\log L_{\rm kin}\,
L_{\rm kin}\,\Phi(L_{\rm kin},z),
\end{equation}
where \(\Phi(L_{\rm kin},z)\) is the kinetic LF obtained from the radio LF via Equation (\ref{eq:klf}), and \(\epsilon_{\rm rad}\) and \(\epsilon_{\rm kin}\) are the radiative and kinetic efficiencies, respectively \citep{1982MNRAS.200..115S,2008ApJ...676..131S}.

In practice, the integral is evaluated over the finite luminosity range, as commonly adopted in radio AGN studies \citep[e.g.,][]{2016MNRAS.460....2P,2017A&A...602A...6S}. Specifically, for the fiducial BHAD calculation we first impose radio-luminosity limits and then convert them to the corresponding kinetic-luminosity limits using Equation~(\ref{eq:Lkin_Lrad}). The lower limit is set to 
\(L_{1.4,\min}=10^{-3}L_\star\) for the fiducial Model C. With the best-fitting 
\(\log(L_\star/{\rm W\,Hz^{-1}})=24.024\), this gives \(\log(L_{1.4,\min}/{\rm W\,Hz^{-1}})=21.024\). The upper limit is set to \(\log(L_{1.4,\max}/{\rm W\,Hz^{-1}})=30\). These limits are adopted at all redshifts in the fiducial calculation.
The dominant contribution is expected to come from the luminosity range where the modeled LF is non-negligible, in particular around the characteristic knee luminosity.
As a robustness check, we also tested a much broader numerical kinetic-luminosity interval and found that the integrated result changes only modestly. This indicates that the BHAD estimate is not driven by the extreme faint-luminosity tail.
In this expression, the last term represents the total mass accreted onto the central black hole for a source of kinetic luminosity $L_{\rm kin}$, i.e., the individual black hole accretion rate
\begin{equation}
\label{eq:bhar_single}
\dot{M}_{\rm BH} = \frac{(1-\epsilon_{\rm rad}-\epsilon_{\rm kin})L_{\rm kin}}{\epsilon_{\rm kin}c^2};
\end{equation}
see \citet{2008ApJ...676..131S} for a detailed derivation.

We adopt \(\epsilon_{\rm rad}=0.1\) and \(\epsilon_{\rm kin}=0.05\) in Equation~(\ref{eq:bhad_kin}). The radiative efficiency \(\epsilon_{\rm rad}=0.1\) is a commonly used fiducial value in So\l{}tan-type arguments that connect the time-integrated AGN emissivity to the local black-hole mass density \citep{1982MNRAS.200..115S,Hopkins2007,2008ApJ...676..131S,Delvecchio2014}. Here the kinetic efficiency is defined as the jet-production efficiency, $\epsilon_{\rm kin}\equiv P_{\rm jet}/(\dot{M}_{\rm acc}c^2)$. We adopt $\epsilon_{\rm kin}=0.05$ as a fiducial few-per-cent efficiency for the radio-loud kinetic channel considered here. This value lies within the broad empirical range inferred for radio galaxies; for example, \citet{2020ApJ...892..116W} found $\eta_{\rm jet}=P_j/(\dot{M}_{\rm acc}c^2)$ spanning from \(\lesssim10^{-3}\) to \(\sim0.2\) in young radio galaxies. We emphasize that the absolute normalization of the inferred BHAD scales as \((1-\epsilon_{\rm rad}-\epsilon_{\rm kin})/\epsilon_{\rm kin}\), and is therefore efficiency-dependent. The resulting BHAD evolution from our fiducial Model~C is shown in Figure~\ref{fig:bhar} and discussed below.

As an additional consistency check, we integrate the inferred BHAD over cosmic time to estimate the corresponding local black-hole mass density associated with the steep-spectrum radio-loud, kinetic-accretion channel:
\begin{equation}
\rho_{\rm BH,0}
=
\int_0^5
\Psi_{\rm bhar}(z)
\left|\frac{dt}{dz}\right|
dz .
\end{equation}

Using the fiducial luminosity limits described above, and adopting
\(\epsilon_{\rm rad}=0.1\) and \(\epsilon_{\rm kin}=0.05\), we obtain
$\rho_{\rm BH,0}=1.281\times10^4\,M_\odot\,{\rm Mpc}^{-3},$
with a statistical \(3\sigma\) range of $[1.257,1.305]\times10^4\,M_\odot\,{\rm Mpc}^{-3}$.
This value is the efficiency-scaled black-hole mass density associated with the steep-spectrum radio-loud kinetic channel, inferred from the kinetic luminosity density using Equation~(\ref{eq:bhad_kin}). 
Compared with the commonly quoted total local SMBH mass density of
\((4\)--\(5)\times10^5\,M_\odot\,{\rm Mpc}^{-3}\)
\citep[e.g.,][]{2004MNRAS.351..169M,2007MNRAS.378..198G,2009ApJ...690...20S}, 
our inferred cumulative contribution corresponds to 2.6\%--3.2\%.
Thus, the time-integrated BHAD inferred here should be interpreted as an efficiency-dependent estimate of the black-hole growth associated with the steep-spectrum radio-loud kinetic channel, rather than as the total black-hole mass density assembled by all AGN accretion modes.

We stress, however, that the quoted statistical interval reflects only the propagated uncertainty from the MCMC posterior under our adopted assumptions, and does not include systematic uncertainties associated with the radio-to-kinetic conversion, efficiency parameters, the adopted radio-luminosity integration limits, or the exclusion of the flat-spectrum radio-loud population.

\subsection{Co-evolution of BHAD and SFRD}

To place the inferred BHAD evolution in a broader galaxy-evolution context, we compare it with the cosmic SFRD. We do not re-derive the SFRD in the present paper. Instead, we directly adopt the radio-based SFRD reconstruction from our previous work (W24, Model C), where the SFRD was obtained from the best-fit SFG RLF by converting radio luminosity into SFR and integrating the analytical LF over luminosity. In the present work, this previously derived SFRD is used only as a comparison curve for the BHAD. This ensures that both the SFRD and BHAD comparisons are anchored to the same LF functional family. 

To enable a meaningful comparison with the BHAD, we rescaled the SFRD by a factor of $0.8 \times 10^{-3}$, following the convention adopted in \citet{Heckman2014}, who noted that the ratio between star formation and black hole accretion has remained approximately constant over cosmic time, with a value of order $10^{3}$.
The comparison is illustrated in Figure~\ref{fig:bhar}, where the blue and red solid lines represent the evolution of the BHAD and the rescaled SFRD, respectively. The shaded regions show the associated $3\sigma$ uncertainties. 
A remarkable similarity is observed between the two evolutionary tracks: both the BHAD and the rescaled SFRD exhibit a pronounced rise from the local Universe toward higher redshift, peaking around $z\sim2$, an epoch commonly referred to as the ``cosmic noon''. We emphasize, however, that the BHAD estimated here is not the total cosmic BHAD, but rather the channel-specific contribution traced by the steep-spectrum radio-loud AGN population. Accordingly, the comparison with the SFRD should be interpreted primarily as a population-level, shape-based comparison of redshift evolution, rather than as a correspondence in absolute normalization or in the total mass-growth budget. Within this restricted interpretation, the similar redshift dependence of the two curves provides a suggestive physical indication that the kinetic-mode growth traced by the steep-spectrum radio-loud AGN sub-population is broadly synchronized, at the population level, with the global star-formation history of galaxies. This does not imply that radio-loud AGNs dominate the total SMBH growth budget, nor does it establish a direct galaxy-by-galaxy causal connection between star formation and radio-mode accretion. Instead, the similarity suggests that both processes may be linked to common cosmological drivers, such as the evolving gas supply, halo growth, and feedback-regulated galaxy evolutio.

A further advantage of our approach is methodological consistency: the SFG LF models used in W24 (Models~B and~C) share the same functional form as the AGN LF models adopted here, differing only in their best-fitting parameters for each population. Thus, a single parametric LF framework can simultaneously describe two physically distinct components—SFGs and radio-loud AGNs—over a wide redshift range. This remarkable agreement underscores the robustness and versatility of our LF framework, and may further imply that the cosmic evolution of star formation and black hole growth are governed by common underlying processes \citep[also see][]{Heckman2014,2014ARA&A..52..415M}, possibly regulated by large-scale gas accretion and feedback mechanisms \citep{Croton2006,2006ApJS..163....1H}.

\section{Conclusions}
\label{sec_conclusion}

We have presented a comprehensive study of the cosmic evolution of radio-loud  active galactic nuclei (AGNs) using a beaming-minimized sample of 4,555 steep-spectrum sources. By adopting a luminosity and density evolution (LADE) framework that separates luminosity evolution (LE) and density evolution (DE), we derive a unified and physically transparent description of the AGN radio luminosity function (RLF) at 1.4\,GHz over $0<z\lesssim5$. Our main conclusions are as follows:

\begin{enumerate}
\item \textit{A LADE description is strongly supported by the data.}
Following the LADE framework of \citet{yuan2017mixture}, we find that the cosmic evolution of steep-spectrum radio-loud AGNs can be well described by a separable combination of DE and LE applied to a flexible local LF family. The best-fitting models reproduce the binned RLFs and are simultaneously consistent with external constraints from the local RLFs and Euclidean-normalized source counts, yielding a globally self-consistent description across the full dynamic range of our composite sample. In our fiducial Model~C, the LE term increases toward cosmic noon ($z\sim2$--3) and then flattens or mildly declines at higher redshift, whereas the DE term decreases monotonically with redshift over the range probed.

\item \textit{The observed luminosity-dependent peak redshift is naturally captured by the LADE framework.}
Our LADE (DE+LE) model reproduces the luminosity-dependent turnover (or peak) redshift $z_{\rm peak}(L)$ inferred from the data---the phenomenology often referred to as ``cosmic downsizing.'' Importantly, within this framework there is no need to impose \emph{a priori} different evolutionary laws for low- and high-power radio-loud AGNs; the apparent luminosity dependence arises self-consistently from the same underlying DE and LE terms acting on an LF whose slope changes with luminosity.

\item \textit{A unified functional architecture describes both star-forming galaxies (SFGs) and AGNs.} A key result is that the same LADE functional family previously calibrated for SFGs in W24 also provides an excellent description of radio-loud AGNs when fitted independently. This a posteriori convergence enables controlled, like-for-like comparisons between the two physically distinct populations within a standardized parametric framework.

\item \textit{The unified two-component model is globally self-consistent.} Combining the SFG component from W24 (Model~C) with our AGN component (Model~C) yields a total 1.4\,GHz RLF and source counts that agree well with the corresponding observational benchmarks, supporting the internal consistency of the two-component decomposition at both the LF and source-count level.

\item \textit{The black-hole growth traced by the steep-spectrum radio-loud AGN population accounts for only a modest fraction of the total local SMBH mass density.}
Using our fiducial radio-to-kinetic conversion, luminosity integration limits, and efficiencies
\((\epsilon_{\rm rad}=0.1\) and \(\epsilon_{\rm kin}=0.05)\), we obtain an efficiency-scaled integrated mass density of
\(\rho_{\rm BH,0}=1.281\times10^4\,M_\odot\,{\rm Mpc}^{-3}\), corresponding to only \(\sim2.6\)--\(3.2\%\) of the commonly quoted local SMBH mass density of
\((4\)--\(5)\times10^5\,M_\odot\,{\rm Mpc}^{-3}\).
Nevertheless, the inferred BHAD for this population shows a redshift dependence similar to the radio-based SFRD from W24, with both histories rising toward and peaking near \(z\sim2\). In this shape-based sense, the similarity suggests that the growth of this radio-loud AGN sub-population is broadly synchronized, at the population level, with the cosmic star-formation history of galaxies.

\end{enumerate}

\begin{acknowledgments}
We thank the anonymous reviewer for the many constructive comments and suggestions, leading to a clearer description of these results. We acknowledge financial support from the Science Fund for Distinguished Young Scholars of Hunan Province (Grant No. 2024JJ2040), the National Natural Science Foundation of China (Grant Nos. 12073069, 12075084, 12275080, and 12393813), the Major Basic Research Project of Hunan Province (Grant No. 2024JC0001), and the Innovative Research Group of Hunan Province (Grant No. 2024JJ1006). Z.Y. is supported by the Xiaoxiang Scholars Programme of Hunan Normal University. 
\end{acknowledgments}

\FloatBarrier  

\bibliography{refs}{}
\bibliographystyle{aasjournal}

\FloatBarrier  
\appendix

\begin{figure*}
        \centering
        \includegraphics[width=\textwidth]{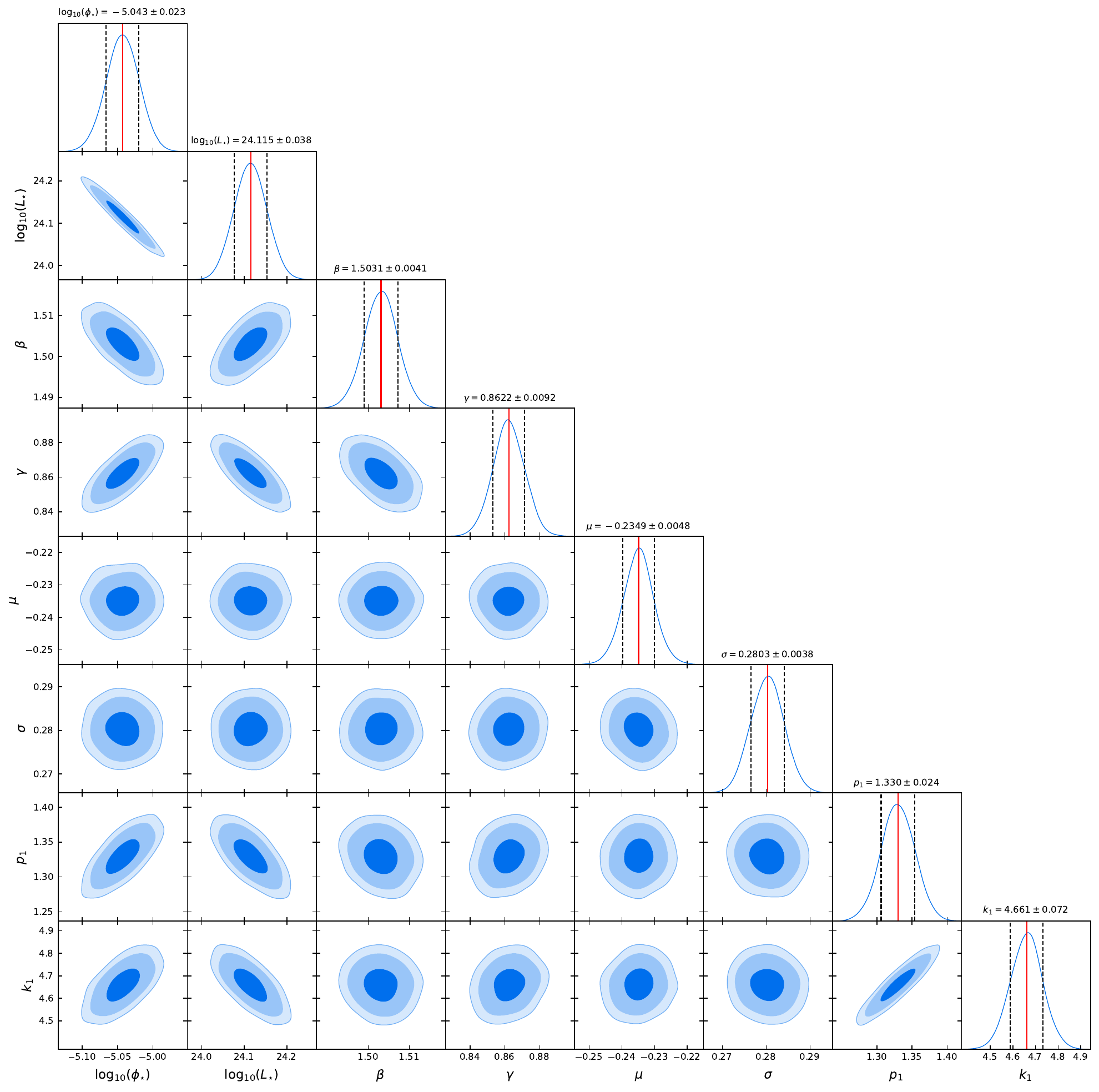}
        \caption{Corner plot illustrating the one- and two-dimensional projections of the posterior probability distributions for Model A, derived from the MCMC sampling. The diagonal panels display the marginalized posterior distributions for each parameter, with the 16th and 84th percentiles indicated by vertical dashed lines. The off-diagonal panels present the two-dimensional joint posterior distributions for each parameter pair, with 1$\sigma$, 2$\sigma$, and 3$\sigma$ confidence levels indicated by blue regions of varying depth. The red vertical solid lines indicate the best-fitting parameter values.}
        \label{fig:cornerplotA}
\end{figure*}

\begin{figure*}
        \centering
        \includegraphics[width=\textwidth]{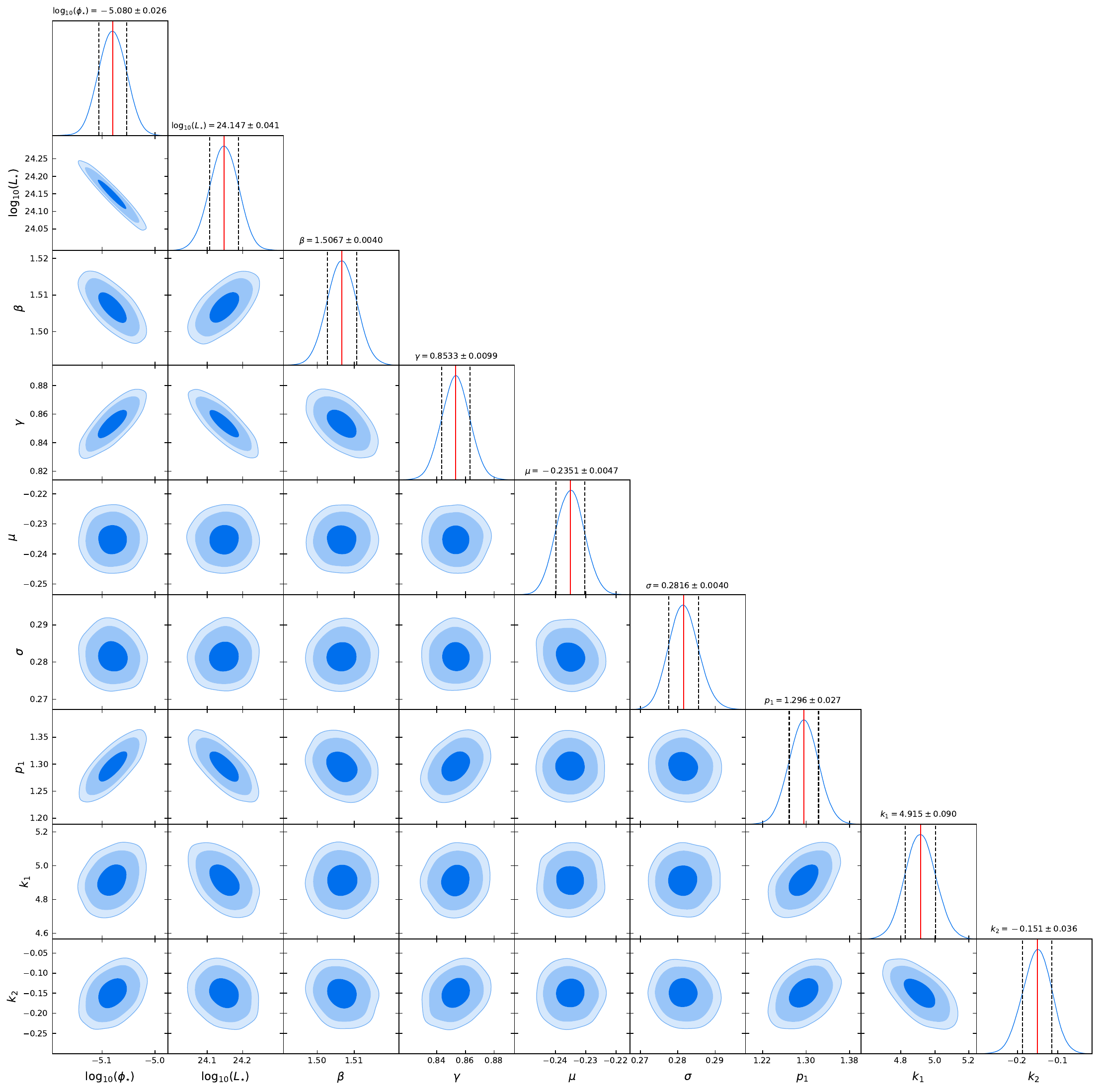}
        \caption{Similar to Figure \ref{fig:cornerplotA}, but for Model B.}
        \label{fig:cornerplotB}
\end{figure*}

\begin{figure*}
        \centering
        \includegraphics[width=\textwidth]{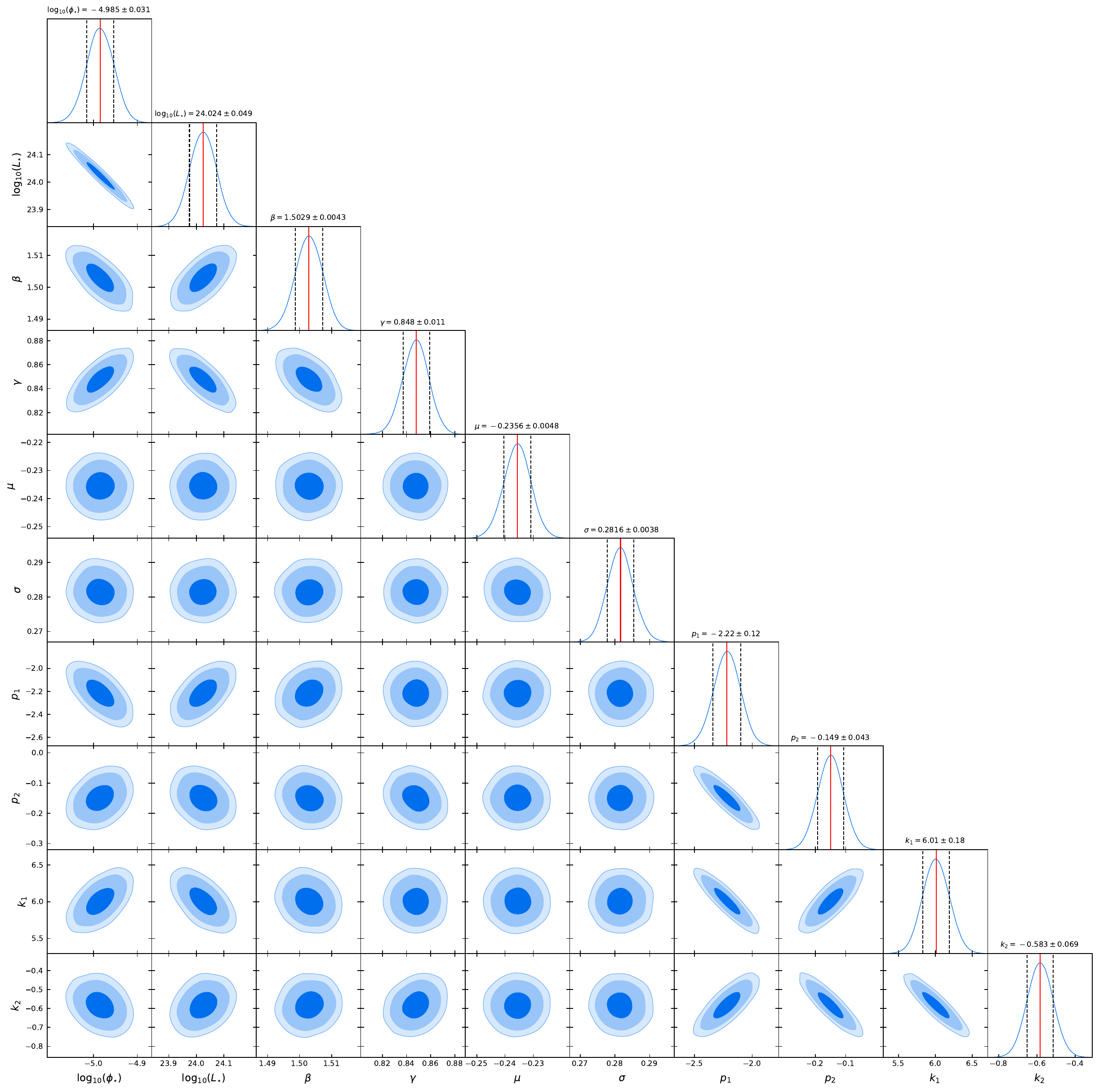}
        \caption{Similar to Figure \ref{fig:cornerplotA}, but for Model C.}
        \label{fig:cornerplotC}
\end{figure*}

\end{document}